\documentclass[
 reprint,
 amsmath,amssymb,
 aps,prb,superscriptaddress
]{revtex4-1}

\usepackage{graphicx}
\usepackage{dcolumn}
\usepackage{bm}
\usepackage{chemfig}
\usepackage{caption}
\usepackage{placeins}
\usepackage{hyperref}
\usepackage{xcolor}
\usepackage{subfig}
\usepackage{siunitx}
\usepackage{ulem}
\usepackage{units}
\usepackage[version=4]{mhchem}
\hypersetup{
    colorlinks=true,
    linkcolor=blue,
    filecolor=magenta,      
    urlcolor=blue,
    citecolor=blue
}

\begin{document}

\title{First-Principles Study of Localised and Delocalised Electronic States in Crystallographic Shear Phases of Niobium Oxide}
\author{Can P. Ko\c{c}er}
\email{cpk27@cam.ac.uk}
\affiliation{Theory of Condensed Matter, Cavendish Laboratory, University of Cambridge, J. J. Thomson Avenue, Cambridge CB3 0HE, UK}
\author{Kent J. Griffith}
\affiliation{Department of Materials Science and Engineering, Northwestern University, Evanston, Illinois, 60208, USA}
\affiliation{Department of Chemistry, University of Cambridge, Lensfield Road, Cambridge CB2 1EW, UK}
\author{Clare P. Grey}
\affiliation{Department of Chemistry, University of Cambridge, Lensfield Road, Cambridge CB2 1EW, UK}
\author{Andrew J. Morris}
\email{a.j.morris.1@bham.ac.uk}
\affiliation{School of Metallurgy and Materials, University of Birmingham, Edgbaston, Birmingham B15 2TT, UK}
\date{\today}

\begin{abstract}
Crystallographic shear phases of niobium oxide form an interesting family of compounds that have received attention both for their unusual electronic and magnetic properties, as well as their performance as intercalation electrode materials for lithium-ion batteries. Here, we present a first-principles density-functional theory study of the electronic structure and magnetism of H-\ce{Nb2O5}, \ce{Nb25O62}, \ce{Nb47O116}, \ce{Nb22O54}, and \ce{Nb12O29}. These compounds feature blocks of niobium-oxygen octahedra as structural units, and we show that this block structure leads to a coexistence of flat and dispersive energy bands, corresponding to localised and delocalised electronic states. Electrons localise in orbitals spanning multiple niobium sites in the plane of the blocks. Localised and delocalised electronic states are both effectively one-dimensional and are partitioned between different types of niobium sites. Flat bands associated with localised electrons are present even at the GGA level, but a correct description of the localisation requires the use of GGA+U or hybrid functionals. We discuss the experimentally observed electrical and magnetic properties of niobium suboxides in light of our results, and argue that their behaviour is similar to that of $n$-doped semiconductors, but with a limited capacity for localised electrons. When a threshold of one electron per block is exceeded, metallic electrons are added to existing localised electrons. We propose that this behaviour of shear phases is general for any type of $n$-doping, and should transfer to doping by alkali metal (lithium) ions during operation of niobium oxide-based battery electrodes. Future directions for theory and experiment on mixed-metal shear phases are suggested.
\end{abstract}

\maketitle

\section{Introduction}
\begin{figure*}
    \centering
    \subfloat[\ce{Nb22O54}]{\includegraphics[scale=0.3]{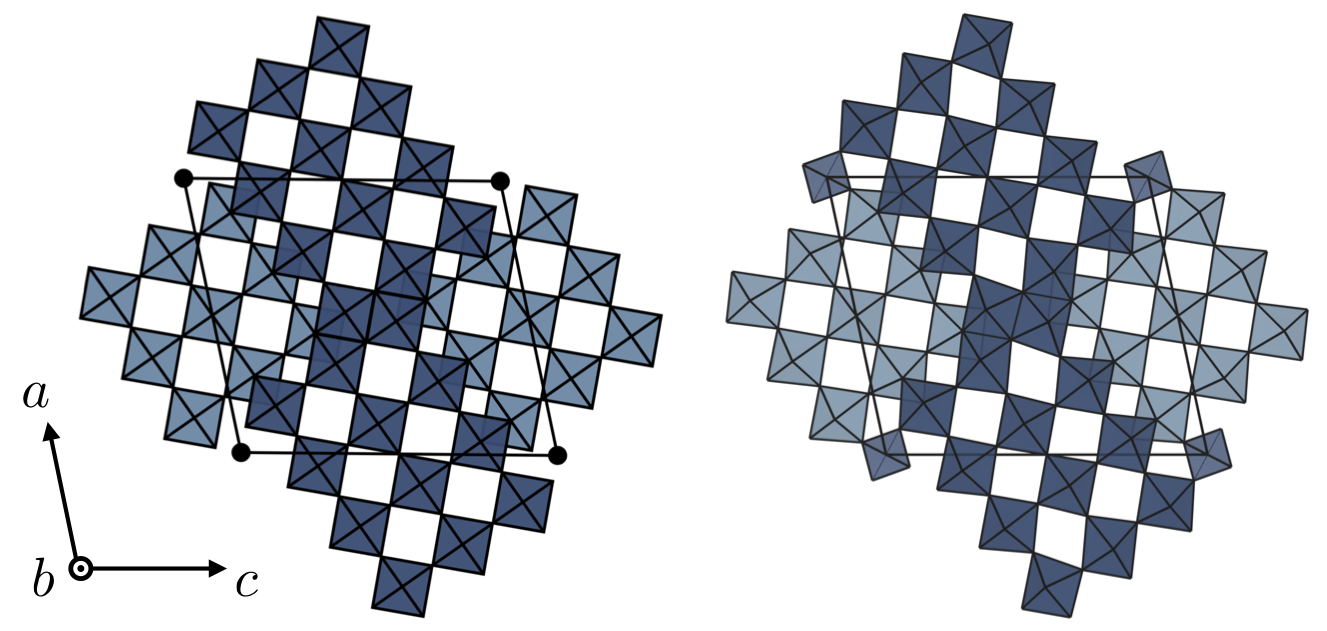}}
    \subfloat[H-\ce{Nb2O5}]{\includegraphics[scale=0.2]{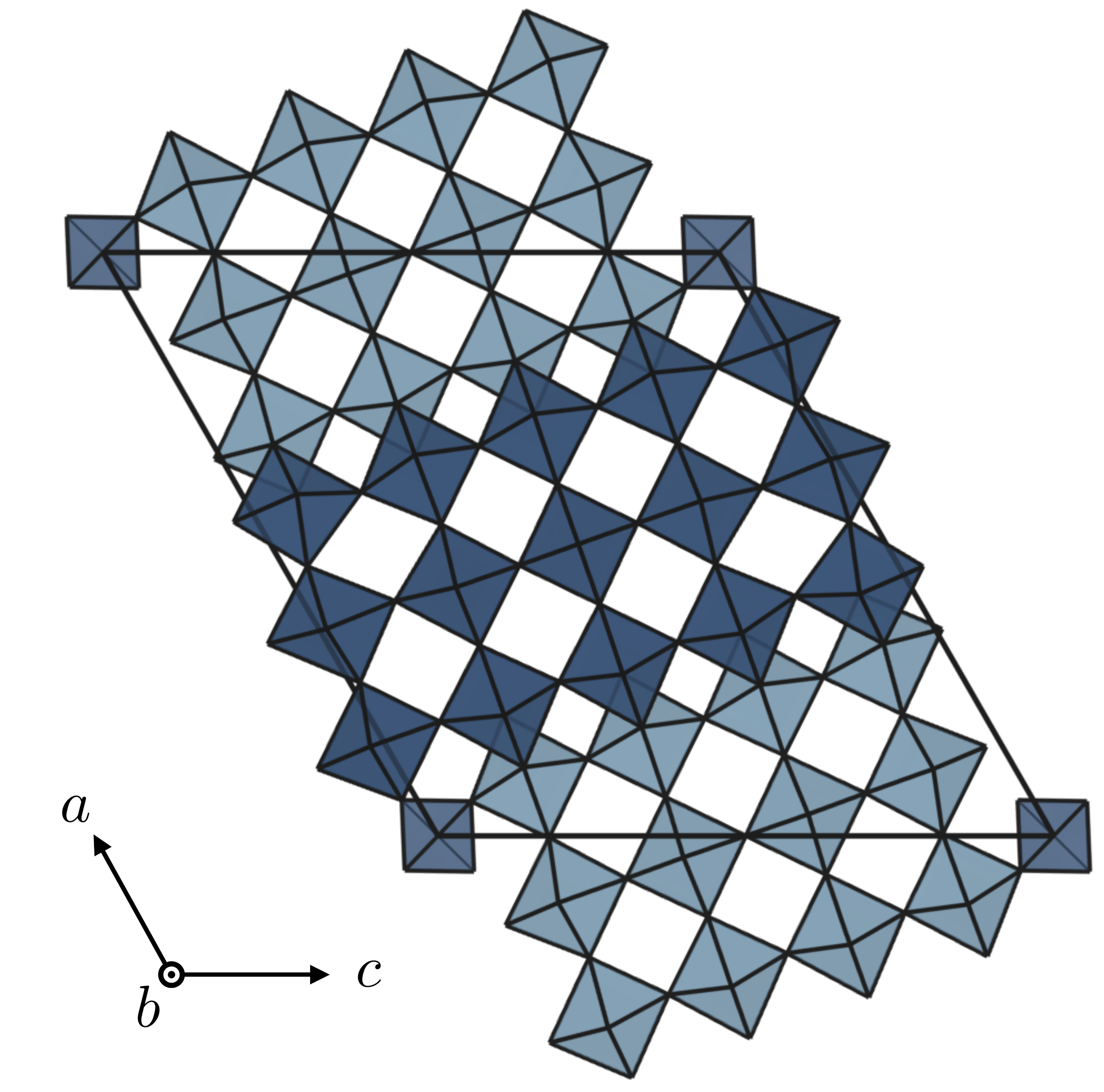}}
    
    \subfloat[{\it m}-\ce{Nb12O29}]{\includegraphics[scale=0.25]{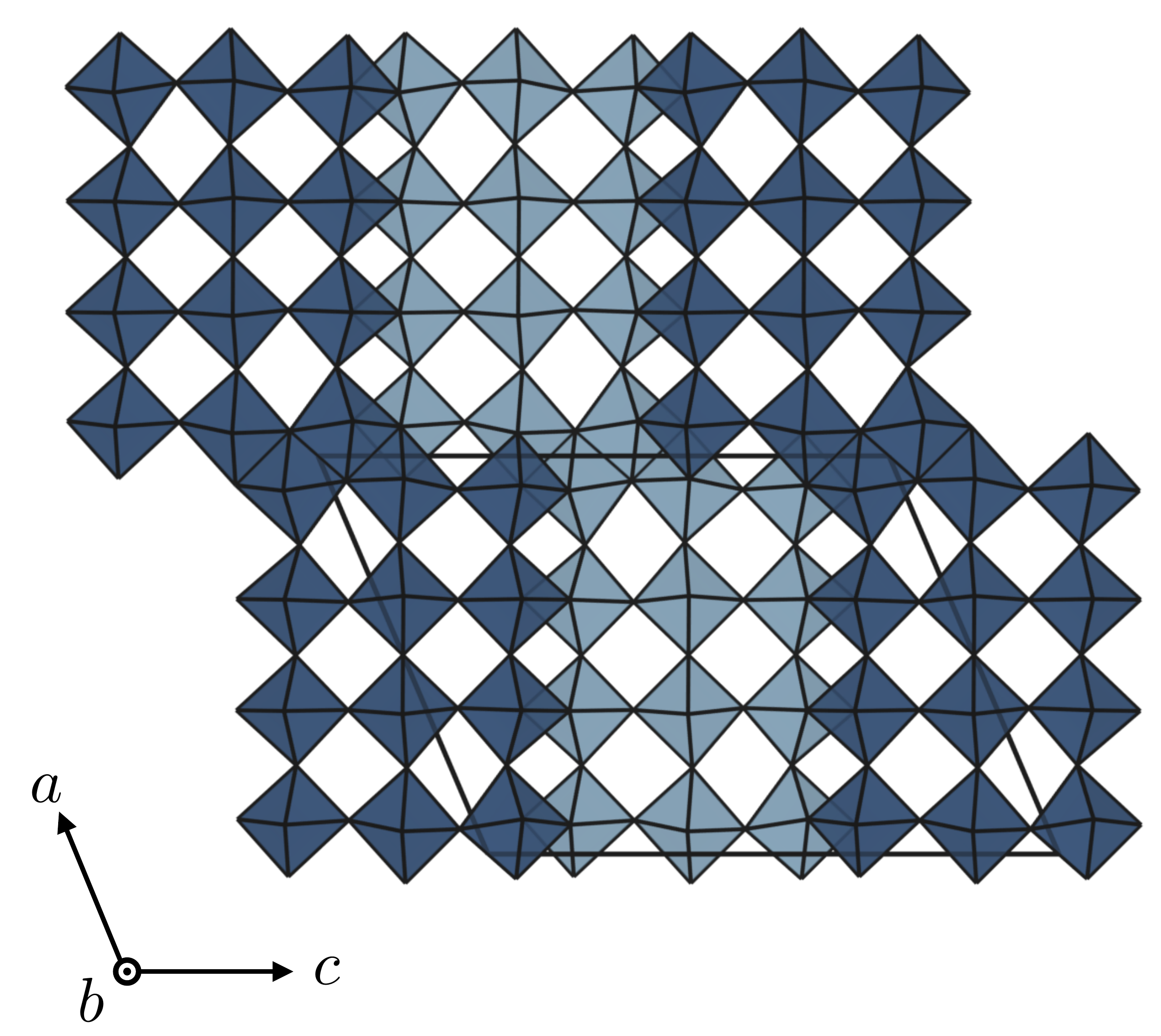}}
    \subfloat[{\it o}-\ce{Nb12O29}]{\includegraphics[scale=0.25]{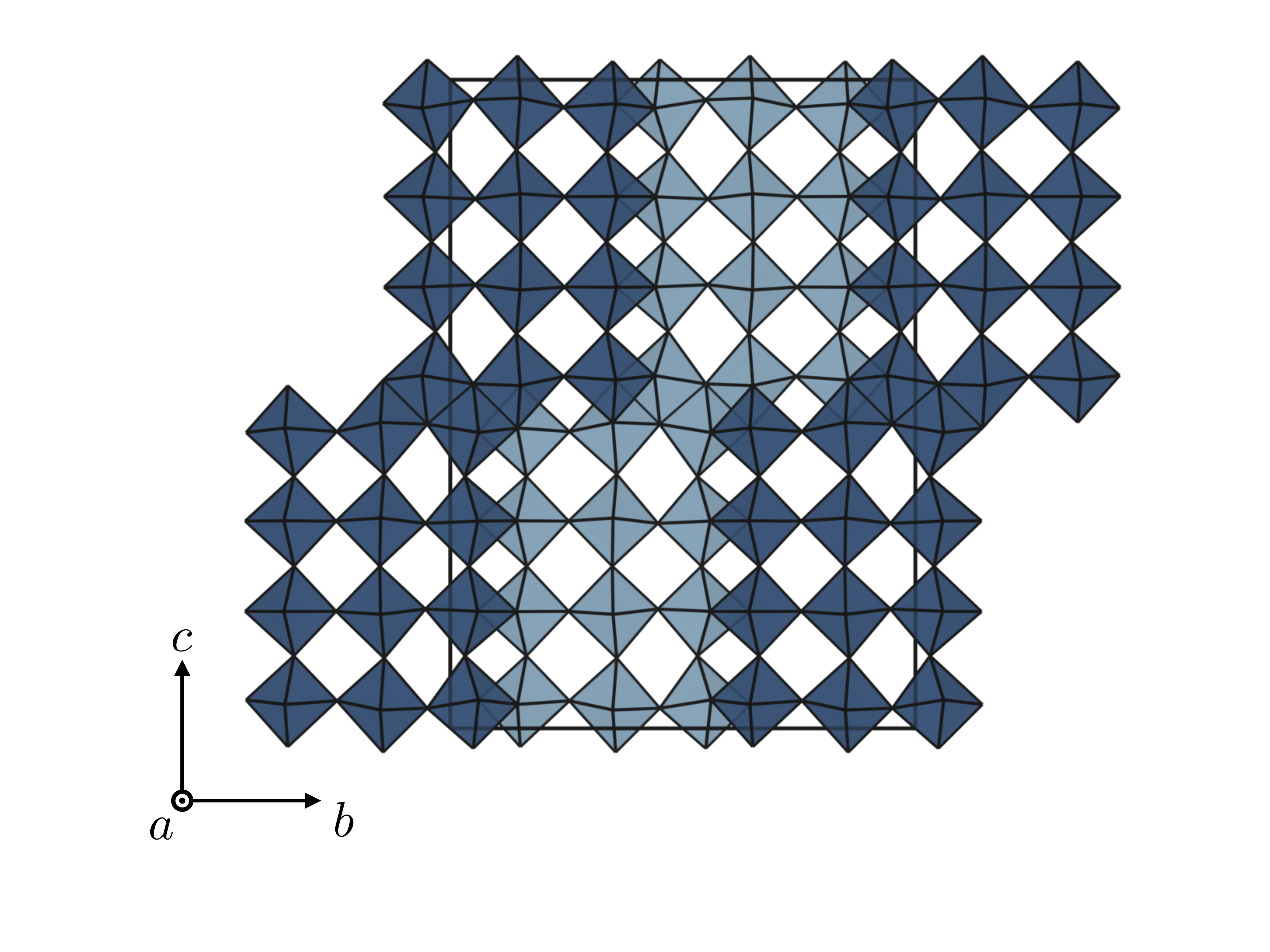}}
    
    \subfloat[\ce{Nb25O62}]{\includegraphics[scale=0.25]{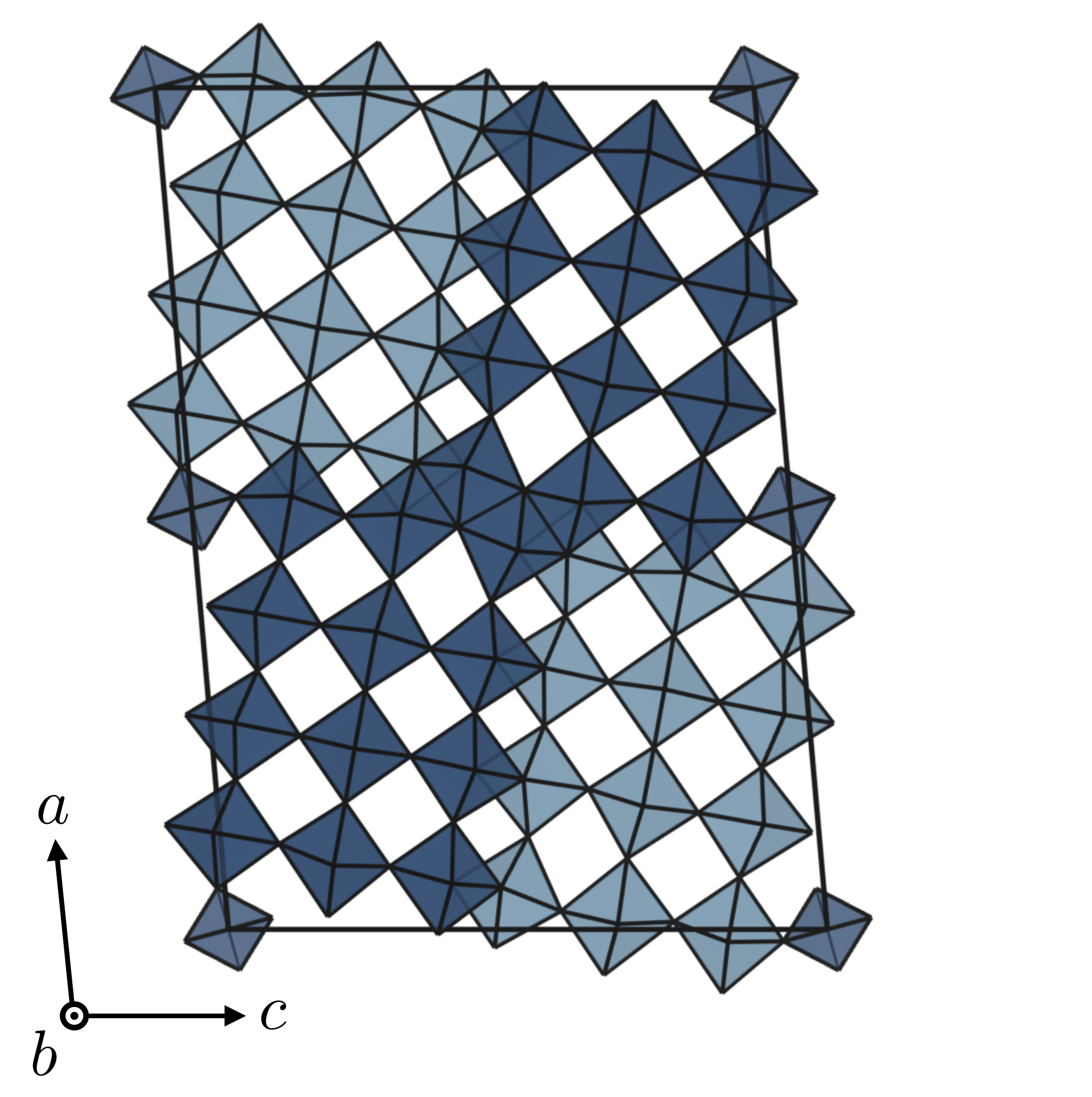}}
    \subfloat[\ce{Nb47O116}]{\includegraphics[scale=0.27]{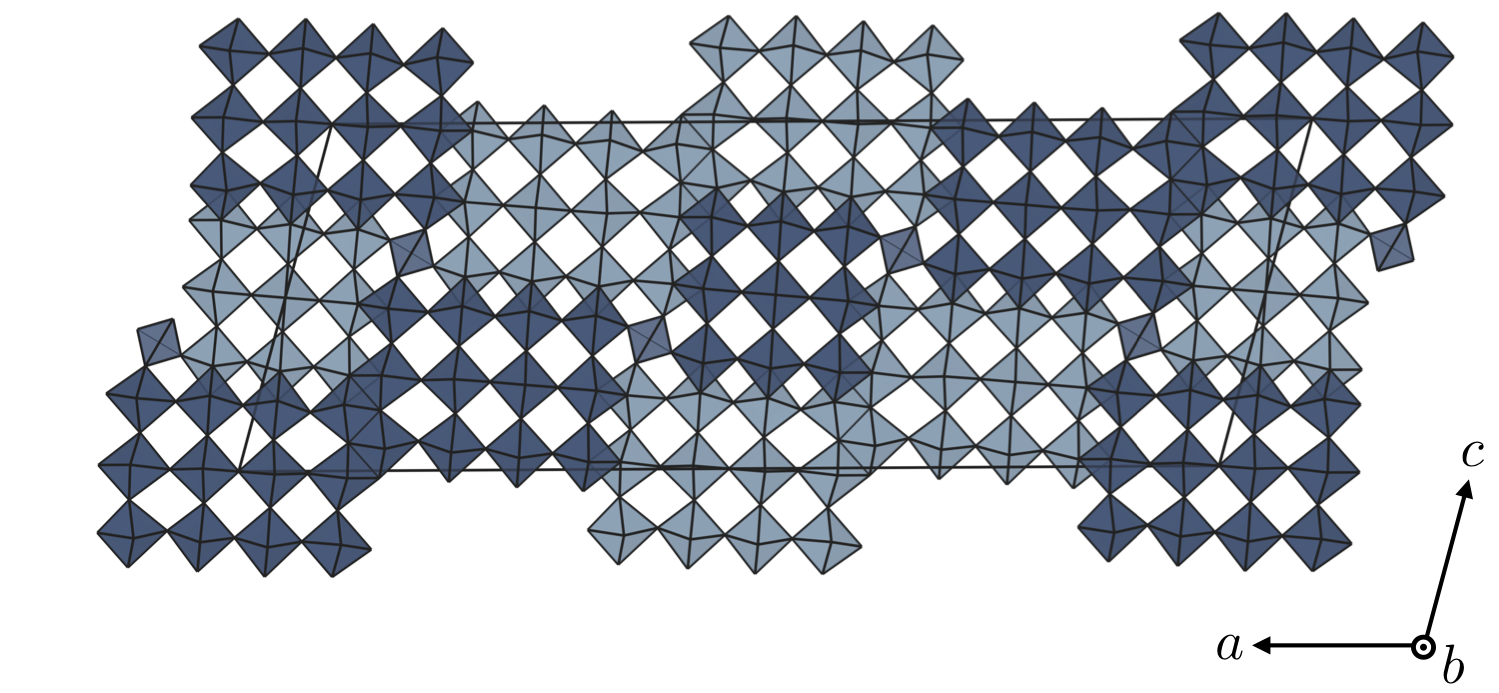}}
    \caption{a) Idealised (left) and locally distorted (right) crystal structure of \ce{Nb22O54}. The tetrahedral site is shown as a black dot in the idealised structure. Crystal structures of b) H-\ce{Nb2O5}, c) monoclinic \ce{Nb12O29}, d) orthorhombic \ce{Nb12O29}, e) \ce{Nb25O62}, and f) \ce{Nb47O116}. Light and dark colored blocks are offset by half of the lattice parameter perpendicular to the plane of the page. Unit cells are outlined in black.}
    \label{fig:xtalstrucs}
\end{figure*}

Transition metal oxides form a fascinating class of compounds with interesting electronic, magnetic, and crystallographic structures. The phase diagram of niobium oxide is especially rich, with a large number of reported phases for \ce{Nb2O5}~\cite{kato1975,kato1976,schafer1966}, in addition to \ce{NbO} and \ce{NbO2}. The high-temperature \ce{Nb2O5} polymorph (H-\ce{Nb2O5}) can be regarded as the parent compound of a family known as crystallographic shear (or Wadsley--Roth) phases~\cite{roth1965a,roth1965d}. In these phases, niobium is present in octahedral coordination, but the Nb/O ratio of \ce{Nb2O5} prevents the formation of purely corner-sharing octahedra. Instead, the structure must include some amount of edge-sharing connections between octahedra. The crystal structures of these compounds consequently consist of {\it blocks} of corner-sharing octahedra of size $n\times m$ that are connected to neighbouring blocks via crystallographic shear planes of edge-sharing connections. In the direction perpendicular to the $n\times m$ plane the units connect infinitely, and tetrahedrally coordinated ions are sometimes present to fill voids in the structure. By reduction of \ce{Nb2O5}, small amounts of Nb$^{4+}$ can be incorporated, and a series of \ce{Nb2O_{5-\delta}} compounds form. These suboxides include \ce{Nb25O62}, \ce{Nb47O116}, \ce{Nb22O54}, and two polymorphs of \ce{Nb12O29} with different crystal symmetries (Fig.~\ref{fig:xtalstrucs}, Table~\ref{tab:strucinfo}). The metal-oxygen octahedra in these compounds are strongly distorted due to a combination of electrostatic repulsion between transition metal ions and the second-order Jahn--Teller effect~\cite{kunz1995,bersuker2006}. Niobium sites in the center of the block are less distorted than those at the periphery. The structural principle of blocks as the building unit, as introduced by Wadsley and Roth, also applies to phases in the \ce{TiO2}-\ce{Nb2O5} and \ce{WO3}-\ce{Nb2O5} phase diagrams~\cite{roth1965d}.

The fully oxidised parent compound \ce{Nb2O5} is a wide bandgap insulator. Low concentrations of valence electrons are introduced through $n$-type doping to form the \ce{Nb2O_{5-$\delta$}} phases. This reduction changes the crystal structure, but the structural motif of the blocks is retained, which makes the niobium suboxides an excellent series of phases to study the interplay between charge state and crystal structure. Magnetic susceptibility measurements show that all \ce{Nb2O_{5-$\delta$}} phases are paramagnetic, with the number of localised moments increasing with $\delta$~\cite{cava1991a,ruscher1991}. Spin interactions are antiferromagnetic and their strength increases with the level of reduction, as indicated by their Curie--Weiss constants. However, only the monoclinic \ce{Nb12O29} phase is found to exhibit long-range antiferromagnetic order, with an ordering temperature of \SI{12}{\kelvin}~\cite{cava1991,andersen2005}. Electrical conductivity measurements show that all \ce{Nb2O_{5-$\delta$}} phases show thermally-activated transport, except for \ce{Nb12O29}, which is metallic down to \SI{2}{\kelvin}~\cite{ruscher1988,cava1991a}. Both electrical and optical measurements indicate that the electron transport in the \ce{Nb2O_{5-$\delta$}} phases is effectively one-dimensional along the block columns~\cite{ruscher1992}. Despite the evidence for localised electrons, single crystal X-ray diffraction studies on \ce{Nb22O54} and {\it o}-\ce{Nb12O29} have not found evidence for charge ordering~\cite{mcqueen2007}. Additional studies have been performed on \ce{Nb12O29} indicating the presence of localised as well as delocalised electrons~\cite{ruscher1988,lappas2002,cheng2009}.

Despite their interesting physical properties, the niobium suboxides have not been investigated with first-principles methods, and the relationship between the level of reduction and the electronic and magnetic properties remains unclear. Two previous first-principles studies have examined the two polymorphs of \ce{Nb12O29}, with rather different conclusions regarding the electronic structure, despite their very strong structural similarity~\cite{fang2014,lee2015}. Additionally, there is new interest in crystallographic shear phases due to their excellent performance as electrode materials in batteries~\cite{griffith2018,li2018}, and the electronic structure of the suboxides is likely to be similar to that of other shear phases.

In this article, we study six different niobium (sub)oxides using density-functional theory calculations and establish common principles governing their electronic and magnetic behaviour. As the 4$d$ band in these materials is highly complex, we first examine \ce{Nb22O54} in detail, and then present results on \ce{Nb12O29}, \ce{Nb47O116}, \ce{Nb25O62}, and H-\ce{Nb2O5}. We show that all of these structures host flat and dispersive bands, which correspond to localised and delocalised electronic states. Each block can host a single localised state in the block plane that is spread over multiple niobium sites. Delocalised states are present along the shear planes. These results are independent of the $n$-type doping,  and alkali metal doped (lithiated) shear phases show similar features to the suboxides, which has implications for their use as battery electrodes. We discuss experimental studies of electrical and magnetic properties of the suboxides in terms of a consecutive filling of localised and delocalised states. Finally, based on these results, we suggest new directions for theory and experiment.

\begin{table*}
    \centering
    \begin{ruledtabular}
    \begin{tabular}{cccccccccc}
        Compound & O/Nb & $N_{e^{-}}/$block & Space group & Source & $a$ & $b$ & $c$ & $\beta$ & Block size(s) \\ \hline
        H-\ce{Nb2O5} & 2.500 & 0 & $P2/m$ & Expt.~\cite{kato1976} & $21.153$ & $3.823$ &  $19.356$ & $119.80^o$ & $3\times 4, 3\times 5$ \\
          & & & $P2$ & PBE & $21.433$ & $3.841$ &  $19.614$ & $119.85^o$ & \\ \hline
        \ce{Nb25O62} & 2.480 & $\nicefrac{1}{2}$ & $C2$ & Expt.~\cite{cava1991a} & 29.78 & 3.825 & 21.14 & $94.7^o$ & $3\times 4$ \\
          & & & & PBE & $30.224$ & $3.84$ &  $21.44$ & $95.0^o$ &  \\ \hline
        \ce{Nb47O116} & 2.468 & $\nicefrac{2}{3}$ & $C2$ & Expt.~\cite{cava1991a} & 57.74 & 3.823 & 21.18 & $105.3^o$ & $3\times 3, 3\times 4$ \\
          & & & & PBE & $58.43$ & $3.84$ &  $21.44$ & $105.3^o$ & \\ \hline
        \ce{Nb22O54} & 2.455 & 1 & $P2/m$ & Expt.~\cite{mcqueen2007} & 15.749 & 3.824 & 17.852 & $102.029^o$ & $3\times 3, 3\times 4$ \\
         & & & $P2$ & PBE & $15.931$ & $3.842$ &  $18.036$ & $102.06^o$ & \\
         & & & $P2$ & PBE+U & $15.935$ & $3.836$ &  $18.061$ & $101.99^o$ & \\ \hline
        {\it m}-\ce{Nb12O29} & 2.417 & 2 & $A2/m$ & Expt.~\cite{waldron2004} & 15.695 & 3.831 & 20.723 & $113.103^o$ & $3\times 4$  \\
         & & & & PBE & $15.903$ & $3.832$ &  $20.967$ & $113.1^o$ & \\
         & & & & PBE+U & $15.885$ & $3.837$ &  $20.950$ & $113.09^o$ & \\ \hline
        {\it o}-\ce{Nb12O29} & 2.417 & 2 & $Cmcm$ & Expt.~\cite{mcqueen2007} & 3.832 & 20.740 & 28.890 & - & $3\times 4$  \\
         & & & & PBE & $3.833$ &  $20.955$ & $29.241$ & - &  \\
         & & & & PBE+U & $3.836$ &  $20.961$ & $29.204$ & - &  \\
    \end{tabular}
    \end{ruledtabular}
    \caption{Structural properties of niobium (sub)oxides. Experimental and DFT optimised lattice parameters $a$, $b$ and $c$ are given in \si{\angstrom}. Structural optimisations with DFT+U were performed with a $U_{\mathrm{eff}}$ value of 4.0 eV on niobium $d$-orbitals. $N_{e^-}$ denotes number of electrons introduced by doping. Difference between experimental and DFT space group choices related to ordering of tetrahedral sites (see text).}
    \label{tab:strucinfo}
\end{table*}

\section{Methods}

All density-functional theory calculations were performed using the planewave DFT code CASTEP~\cite{clark2005} (version 18.1). Pseudopotentials including Nb $4s$, $4p$, $4d$ and $5s$, O $2s$ and $2p$, and Li $1s$ and $2s$ states were used for all calculations. Calculations using hybrid functionals employed norm-conserving pseudopotentials~\cite{hamann1979}, all other calculations were performed using Vanderbilt ultrasoft pseudopotentials~\cite{vanderbilt1990}. Crystal structures were obtained from the Inorganic Crystal Structure Database~\cite{hellenbrandt2004} (ICSD) when available. The structure of \ce{Nb47O116} was constructed as described in Ref.~\onlinecite{cava1991a} as a unit cell intergrowth of \ce{Nb25O62} and \ce{Nb22O54} since no crystallographic data, other than the lattice parameters, was available. The space groups of H-\ce{Nb2O5} and \ce{Nb22O54} are reported as both $P2$ and $P2/m$ in the literature~\cite{kato1976,cava1991a,mcqueen2007}. These two space group choices differ only in the full or partial occupancy of the tetrahedral site. For modelling purposes, the ion on the tetrahedral site has to be ordered, resulting in space group $P2$. Atomic positions and lattice parameters of the structures were relaxed using the gradient-corrected Perdew--Burke--Ernzerhof (PBE) functional~\cite{perdew1996}, until the maximum force on any atom was smaller than $0.01$ eV/$\si{\angstrom}$. The calculations used a planewave kinetic energy cutoff of 800 eV for ultrasoft pseudopotentials, and 900 eV for norm-conserving pseudopotentials, unless otherwise stated. The Brillouin zone was sampled with a Monkhorst--Pack grid~\cite{monkhorst1976} finer than $2\pi \times 0.03$ \si{\angstrom}$^{-1}$. Lattice parameters obtained from the structural relaxations are listed in Table~\ref{tab:strucinfo}, and agree very well with the experimental values. Crystallographic information files (CIF) of the PBE optimised structures are available in the Supplemental Material of this article~\footnote{See Supplemental Material at \url{http://link.aps.org/supplemental/10.1103/PhysRevB.99.075151} for crystallographic information files (CIF) of the structures dealt with in this article.}. All electronic structure calculations were performed for antiferromagnetic spin arrangements in the conventional unit cells, as antiferromagnetic spin interactions are observed experimentally~\cite{cava1991a}.

Semilocal density functionals suffer from self-interaction error, which can be alleviated by the use of DFT+U. For calculations in this work, the DFT+U implementation in CASTEP~\cite{dudarev1998} was used, which defines an effective $U$ value $U_{\mathrm{eff}}=U-J$. A value of $U_{\mathrm{eff}}=4.0$ eV was chosen for the Nb $d$ orbitals, in line with other studies on niobium oxides that employed similar implementations of DFT+U within planewave codes~\cite{nahm2008}. The results presented herein are mostly insensitive to the exact value of the $U_{\mathrm{eff}}$ parameter if it lies in the range 3--5 eV, even though the value of the bandgap does depend on the choice of the $U_{\mathrm{eff}}$ parameter. The structures of \ce{Nb22O54} and the \ce{Nb12O29} polymorphs were additionally optimised with PBE+U, and the results are listed in Table \ref{tab:strucinfo}. PBE and PBE+U lattice parameters agree closely, and PBE+U bandstructure and density of states calculations for compounds other than \ce{Nb22O54} and \ce{Nb12O29} were performed on PBE optimised structures.

Hybrid functionals are another way to correct the self-interaction error of semilocal functionals. The range-separated HSE06 functional~\cite{heyd2003} was used to calculate the bandstructure for \ce{Nb22O54}. Due to the significant additional expense incurred by the use of hybrid functionals, the computational parameters for the calculations of bandstructures at the HSE06 level are coarser. The unit cell of \ce{Nb22O54} contains 610 valence electrons, but since the cell is rather short in one particular dimension and extended in the other two, one cannot use only the $\Gamma$-point in the Brillouin zone (BZ) sampling. Instead, a $1\times 5\times 1$ $\Gamma$-centered $\mathbf{k}$-point grid was used in the HSE06 self-consistent field calculations for \ce{Nb22O54}.

Bandstructure calculations were performed for high-symmetry Brillouin zone directions according to those obtained from the SeeK-path package~\cite{hinuma2017}, which relies on spglib~\cite{togo2018}. A spacing between $\mathbf{k}$-points of $2\pi\times 0.025$ \si{\angstrom}$^{-1}$ was used. Density of states calculations were performed with a grid spacing of $2\pi\times 0.01$ \si{\angstrom}$^{-1}$, and the results were postprocessed with the OptaDOS package~\cite{morris2014a}, using the linear extrapolative scheme~\cite{pickard1999,pickard2000}. The c2x~\cite{rutter2018} utility and VESTA~\cite{momma2011} were used for visualisation of wavefunction and density data. Data analysis and visualisation was performed with the matador~\cite{evans} package.

\section{Results}

\subsection{\texorpdfstring{\ce{Nb22O54}}{ERROR}}

\ce{Nb22O54} crystallises in space group $P2/m$~\cite{mcqueen2007}, and shows an ordered mixture of $3\times 3$ and $3\times 4$ blocks of octahedra, in addition to a tetrahedral site (Figure~\ref{fig:xtalstrucs}a). Assuming an ionic model, the compound can be described as $(\mathrm{Nb}^{5+})_{20}(\mathrm{Nb}^{4+})_2(\mathrm{O}^{2-})_{54}$, with two $4d$ electrons per 22 Nb atoms (1 $e^{-}$ per block, Table~\ref{tab:strucinfo}).

\begin{figure}[!htb]
    \centering
    \includegraphics[scale=0.23]{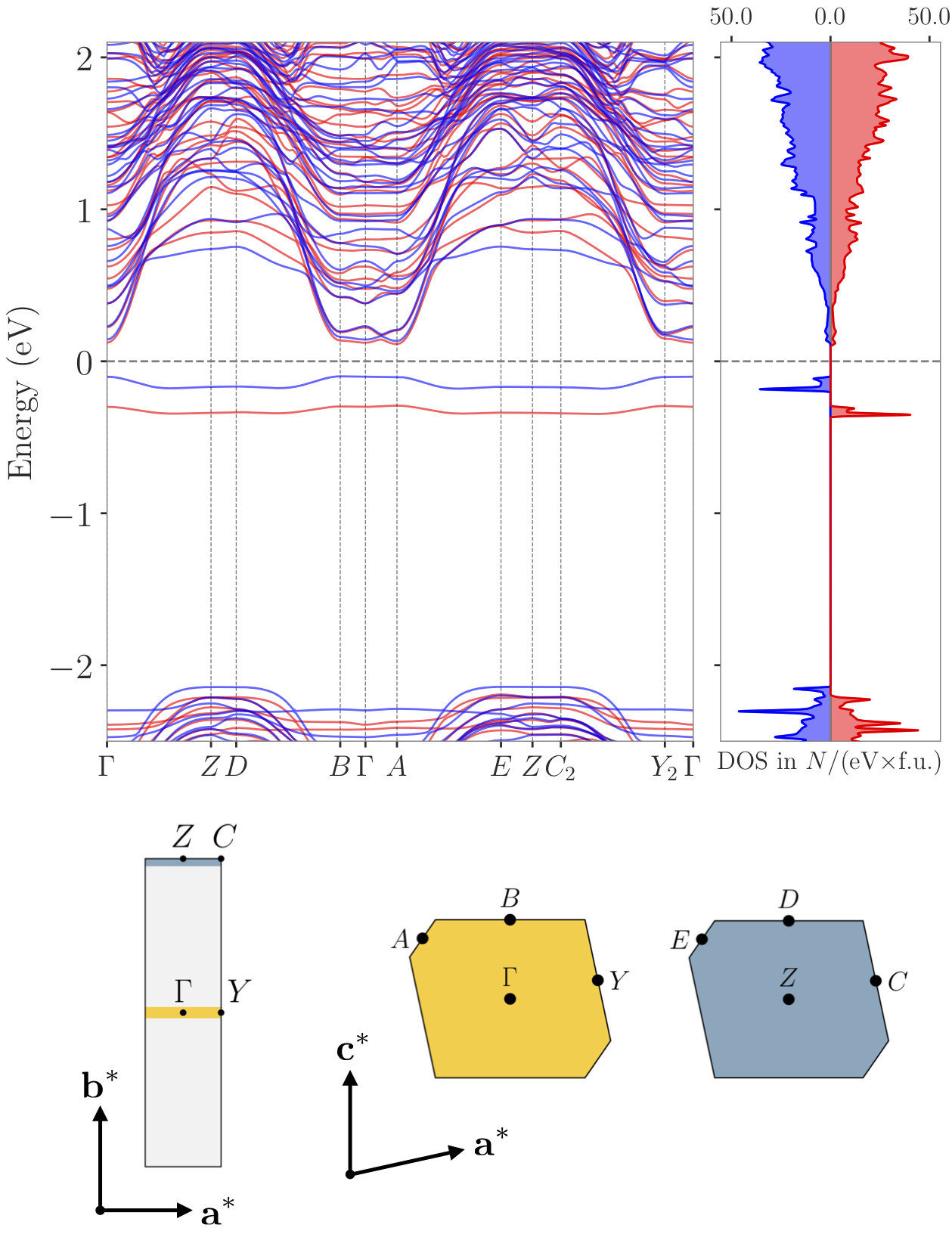}
    \caption{Spin-polarised bandstructure and electronic density of states of \ce{Nb22O54} (PBE+U, $U_{\mathrm{eff}}=4$ eV). Up and down spins colored in red and blue. High symmetry points are marked on slices through the first Brillouin zone. The flat bands below the Fermi level (dashed line) represent localised states.}
    \label{fig:Nb22O54_DFT+U_BS}
\end{figure}

\begin{figure}[!htb]
    \centering
    \subfloat[]{\includegraphics[scale=0.225]{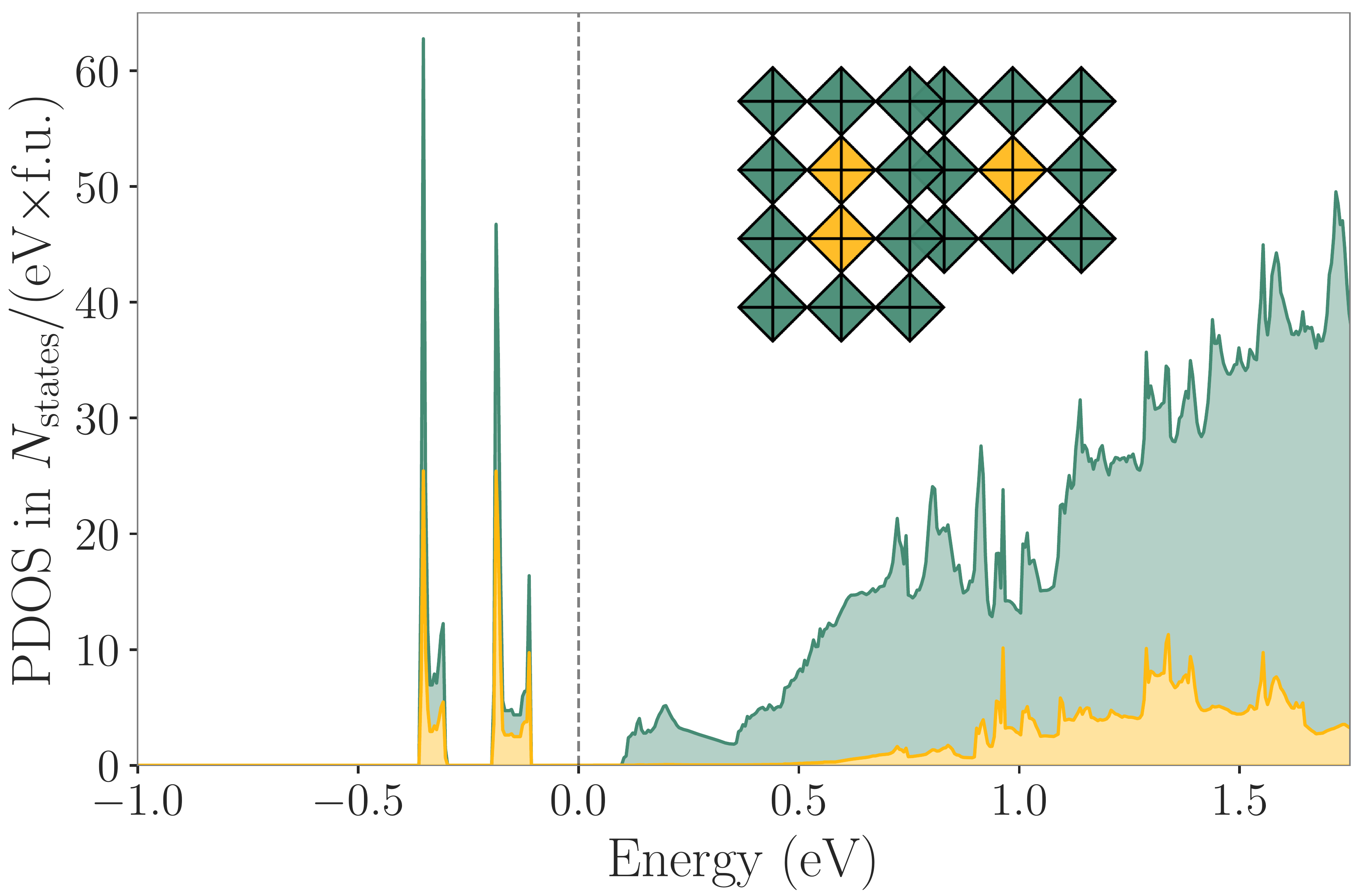}}

    \subfloat[]{\includegraphics[scale=0.225]{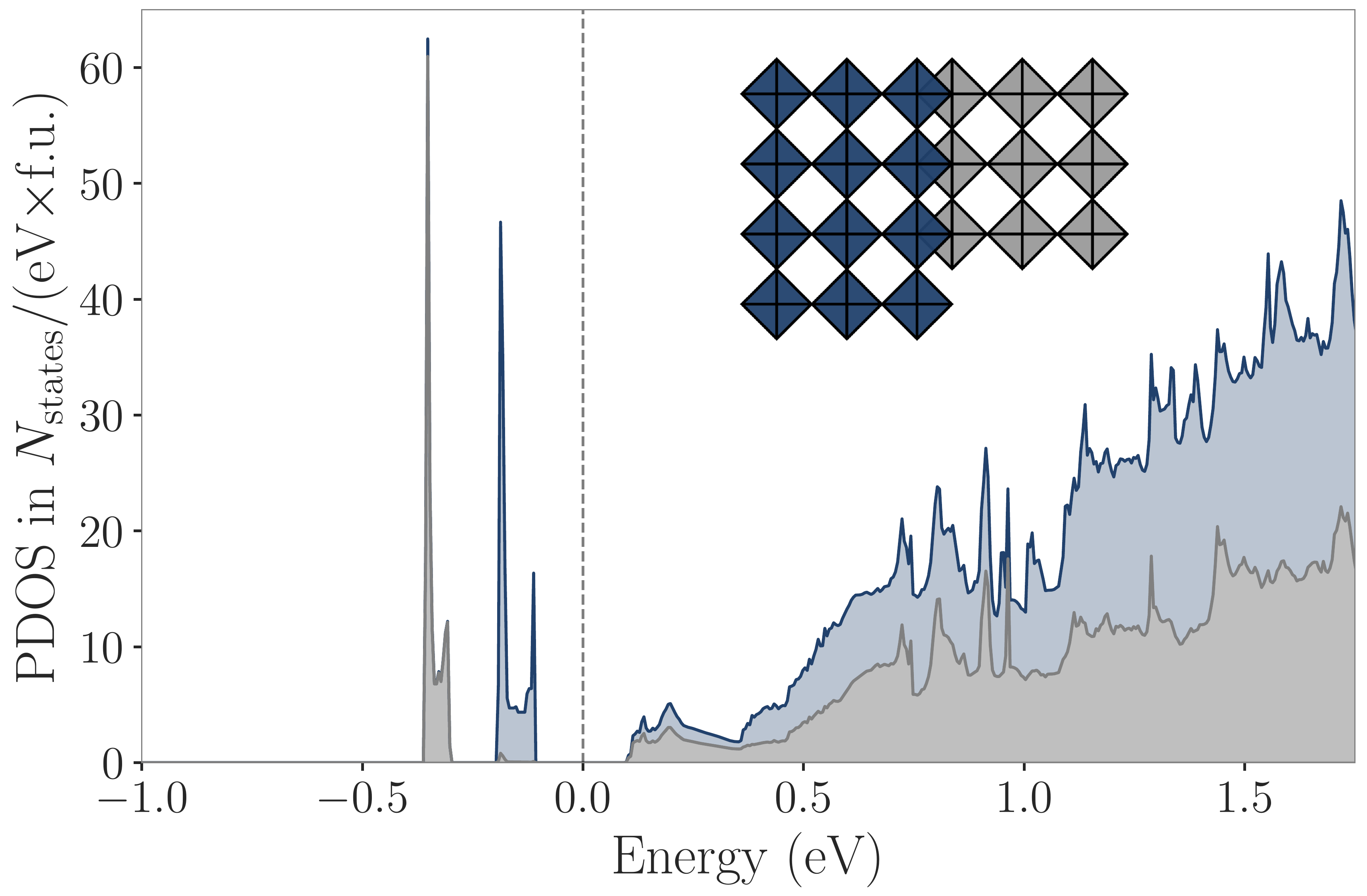}}
    \caption{Spin-summed projected density of states (PBE+U) for \ce{Nb22O54}. Fermi level is indicated by the dashed line. a) PDOS for central (gold) and peripheral (green) niobium sites. b) DOS projected for sites in different blocks, demonstrating separate localisation of electrons in $3\times 3$ and $3\times 4$ blocks. Contributions from sites are proportional to the shaded area.}
    \label{fig:Nb22O54_PDOS}
\end{figure}

The $\mathbf{a}$ and $\mathbf{c}$ lattice vectors of \ce{Nb22O54} are longer than $\mathbf{b}$, which is perpendicular to the block plane (Fig.~\ref{fig:xtalstrucs}, Table~\ref{tab:strucinfo}). The Brillouin zone (BZ) therefore has one long (along $\mathbf{b}^{\ast}$) and two short directions. The PBE+U spin-polarised bandstructure and electronic density of states (DOS) of \ce{Nb22O54} show a large gap between the valence and conduction bands, which are of oxygen $2p$ and niobium $4d$ character, respectively (Fig.~\ref{fig:Nb22O54_DFT+U_BS}). Two fully occupied flat bands (one for each spin) lie within the band gap, leading to the peaks in the DOS below the Fermi level. The flat bands have a very small one-dimensional dispersion, as evidenced by the shapes of the corresponding peaks in the DOS, and represent localised states. In addition to the flat bands, a set of dispersive bands exists just above the Fermi level, which show the largest dispersion along $\mathbf{b}^{\ast}$. The separation between the flat bands and the rest of the conduction states is smallest at special points lying in the $\mathbf{a}^{\ast}$-$\mathbf{c}^{\ast}$ plane of $\Gamma$ ($Y$, $A$, $B$), and largest in the parallel plane at the BZ boundary ($Z$, $C$, $D$, $E$). Due to this pattern, the dispersive bands are also effectively one-dimensional.

With 12 inequivalent niobium sites in the unit cell of \ce{Nb22O54}, site-resolved projected densities of states (PDOS) are complicated and difficult to interpret. More insight is gained by summing PDOS for sets of sites. Figure~\ref{fig:Nb22O54_PDOS}a shows the projection onto different types of niobium sites within the structure, which are classified as central and peripheral, depending on where they sit within the block. We note two things: (1) Both peripheral and central niobium sites contribute to the localised states, even though the contribution of the central sites is greater given the ratio of the two; and (2) only peripheral niobium sites contribute to the unoccupied density of states above the Fermi level (until 0.5 eV above), the contribution from the central sites is exactly zero. The PDOS resolved by block in Fig. \ref{fig:Nb22O54_PDOS}b demonstrates that one localised state is contained in the $3\times 4$ block, and the other, lower energy one, in the $3\times 3$ block. Both blocks contribute roughly equally to the density of unoccupied conduction states.

\begin{figure}[!htb]
    \centering
    \includegraphics[scale=0.25]{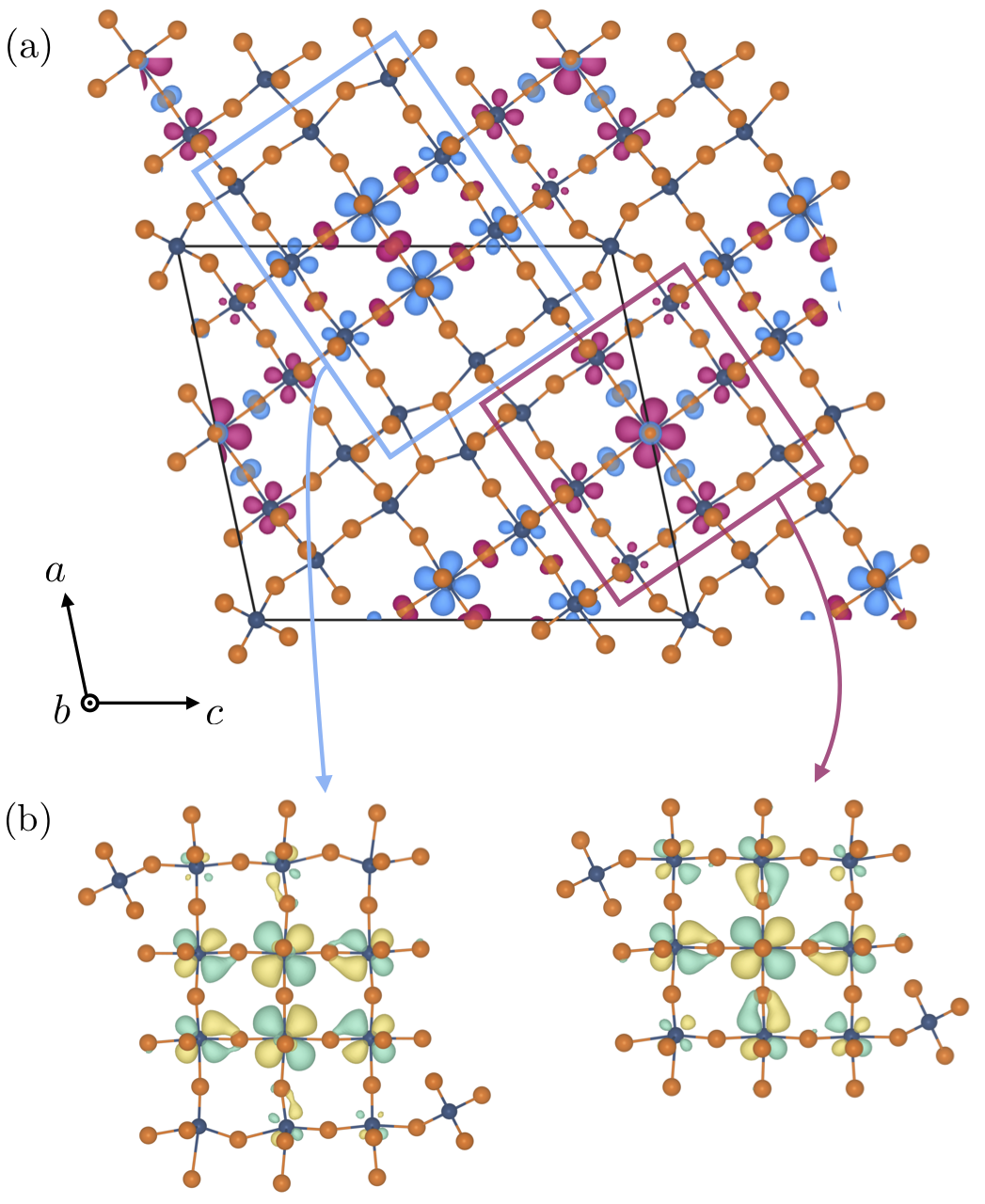}
    \caption{a) Spin density plot of \ce{Nb22O54}. Niobium and oxygen shown in dark blue and orange, respectively. Purple and light blue represent up and down spin density, respectively. The rectangles outline the $3\times 4$ and $3\times 3$ blocks. Spin density isosurface drawn at a value of 0.03 $e^-/\si{\angstrom}^3$. b) Kohn-Sham orbitals associated with localised states (flat bands in Fig. \ref{fig:Nb22O54_DFT+U_BS}) in $3\times4$ and $3\times3$ blocks, different phases of the orbitals shown in yellow and light green.}
    \label{fig:Nb22O54_Spindensity}
\end{figure}

Spin density in \ce{Nb22O54} is predominantly located on the central niobium sites (Fig.~\ref{fig:Nb22O54_Spindensity}a), which also dominate the relevant states as seen from the PDOS (Fig.~\ref{fig:Nb22O54_PDOS}a). One spin is located in each block, and the spin arrangement is antiferromagnetic between the two blocks. However, the ferromagnetic arrangement is only marginally higher in energy (less than 1 meV), indicating very weak spin interactions that are likely a result of the long (nm) magnetic interaction lengths. Kohn--Sham orbitals that are occupied by these localised electrons span the entire block, but only have contribution from niobium sites in the same block (Fig.~\ref{fig:Nb22O54_Spindensity}b). The flat dispersion is a result of the very weak face-on overlap ($\delta$-overlap) between these orbitals along $\mathbf{b}$. Both localised orbitals are similar in appearance, despite the different sizes of the blocks. This suggests that the presence of these states is a general feature of block-type structures.

\begin{figure}[!htb]
    \centering
    \includegraphics[scale=0.4]{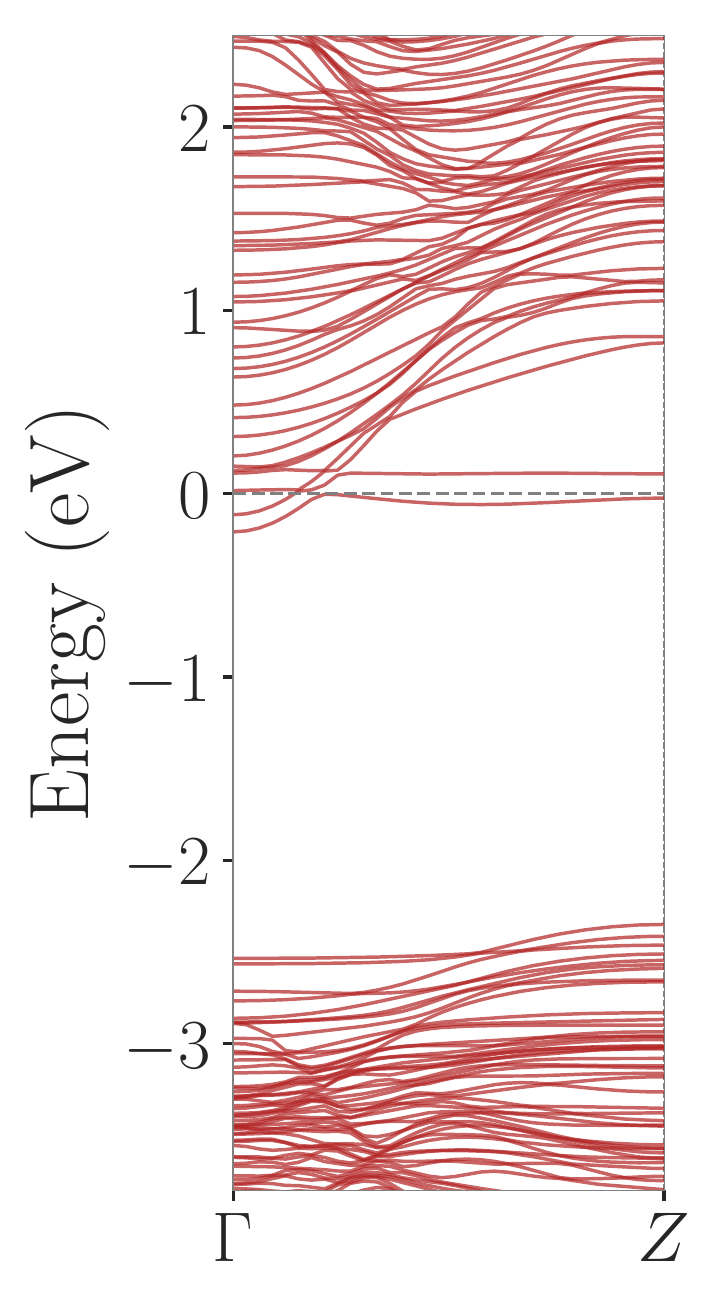}
    \includegraphics[scale=0.4]{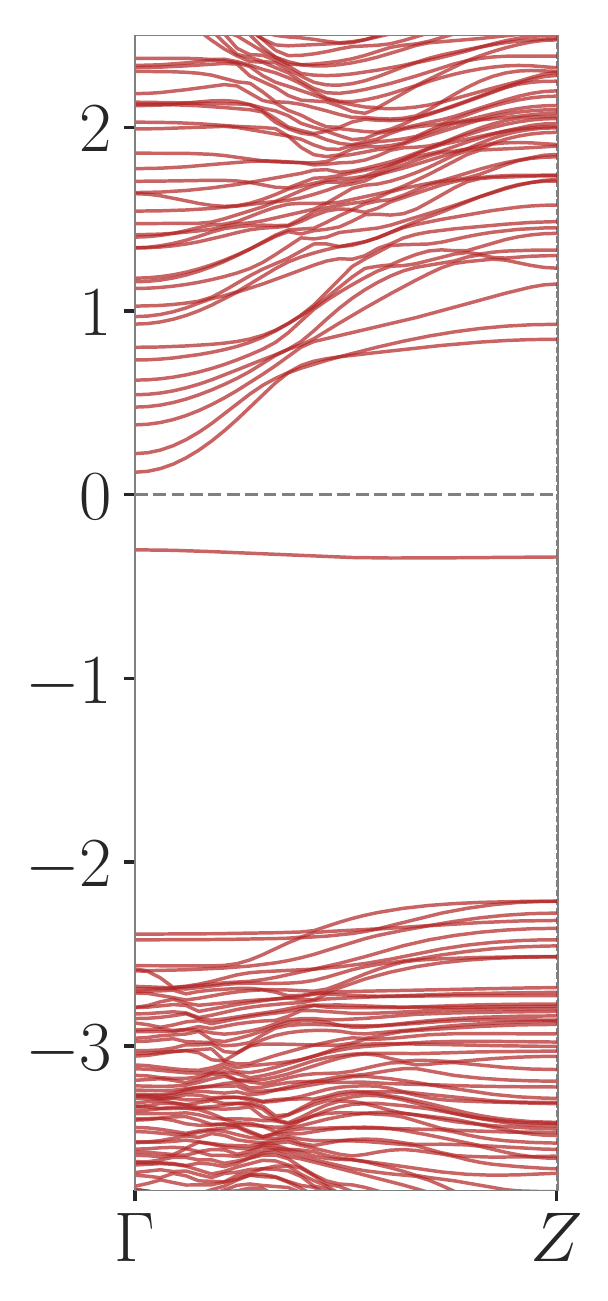}
    \includegraphics[scale=0.4]{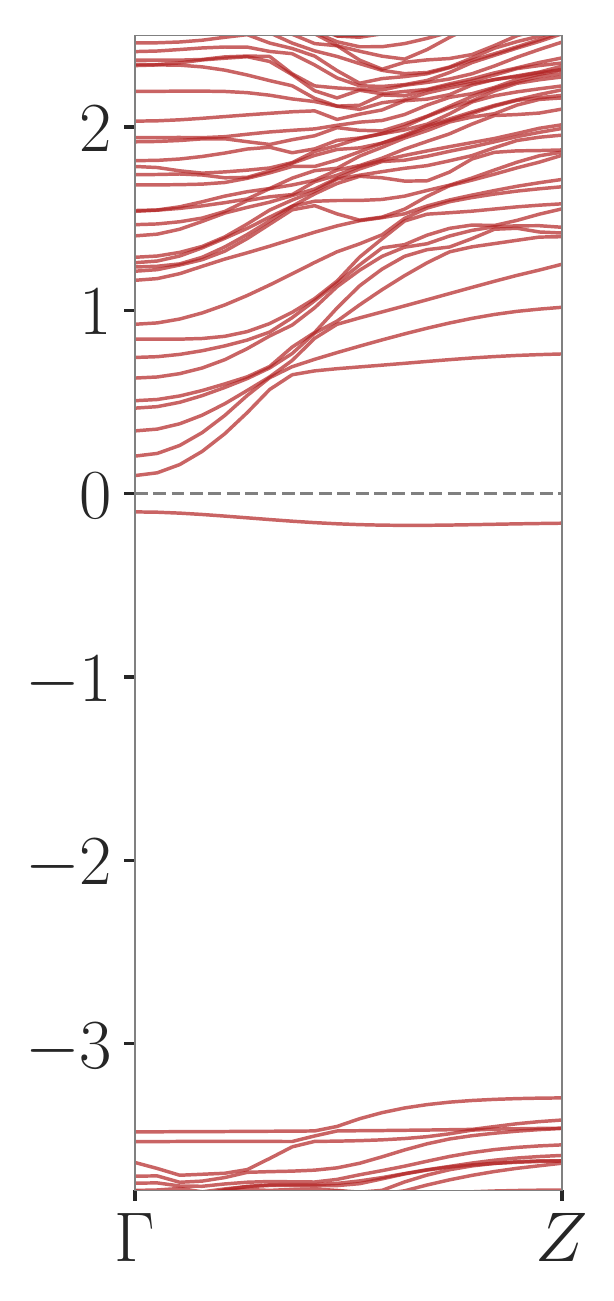}
    \caption{$\Gamma\to Z$ segment of the bandstructure of \ce{Nb22O54} calculated with different levels of theory. PBE, PBE+U ($U_{\mathrm{eff}}=4.0$ eV), and HSE06, from left to right. Only one spin component is shown for clarity.}
    \label{fig:Nb22O54_Theorycomp}
\end{figure}

The results presented above were obtained from PBE+U ($U_{\mathrm{eff}}=4$ eV) calculations. The $\Gamma\to Z$ segment of the \ce{Nb22O54} bandstructure computed with HSE06 and PBE is compared to the PBE+U result in Fig.~\ref{fig:Nb22O54_Theorycomp}. Only the up-spin component is shown, which is associated with the localised electron in the $3\times 3$ block. The bandstructure looks similar for all functionals, and importantly, the relevant feature of localised states, i.e. the flat bands, are present even at the PBE level. However, PBE places the flat bands within the dispersive conduction bands, and both are partially occupied, while both DFT+U and the HSE06 functional place the flat bands below the other conduction bands. PBE also places the opposite spin partner of the localised state in the other block much lower in energy than either PBE+U or HSE06. The precise placement of the flat bands depends on the $U$ value, but in the tested range of 2-5 eV the flat bands are placed below the conduction bands, and the gap between them increases by approximately 200 meV per increase in $U$ by 1 eV. The degree of electron localisation depends on the presence of a gap between flat and dispersive bands. PBE implies metallic behaviour with localised electrons, while HSE06 and PBE+U show full localisation of the electrons. A major difference between the HSE06 and PBE or PBE+U calculations is the size of the gap between valence and conduction bands, which is larger by approximately 1.2 eV for HSE06 compared to PBE+U. The spin density and Kohn-Sham orbitals were plotted from the output of PBE+U calculations, but we note that the results from PBE and HSE06 are visually indistinguishable from the PBE+U results.

\subsection{\texorpdfstring{\ce{Nb12O29}}{Nb12O29}}

\ce{Nb12O29} is more reduced than \ce{Nb22O54} and hosts two $4d$ electrons per 12 niobium sites (i.e. 2 per block, Table~\ref{tab:strucinfo}). The two \ce{Nb12O29} polymorphs are structurally similar, and only differ in the long-range arrangement of the blocks; in the monoclinic polymorph the blocks form a ribbon along $a$, while in the orthorhombic structure the blocks zig-zag along $c$ (Fig.~\ref{fig:xtalstrucs}). 

\begin{figure}[!htb]
    \centering
    \includegraphics[scale=0.52]{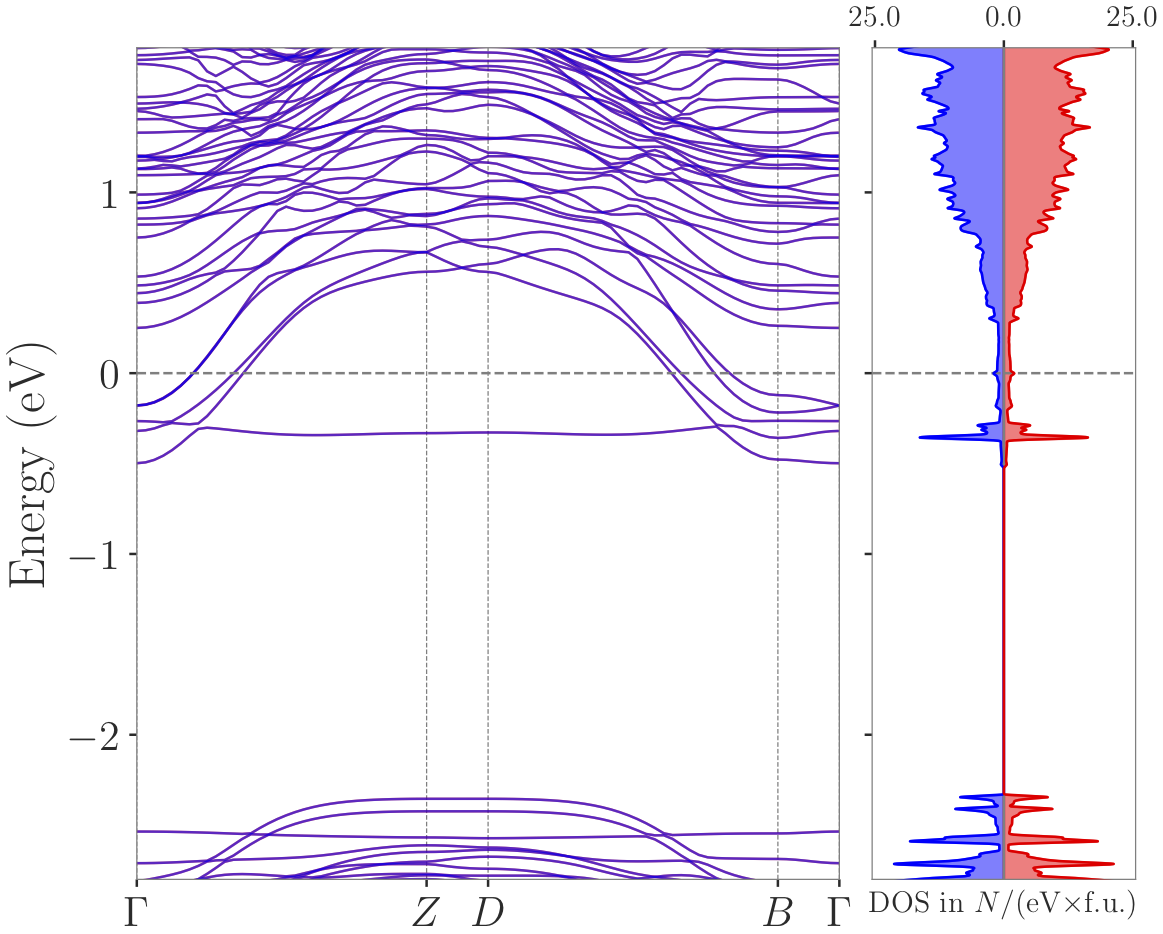}
    \caption{Bandstructure and density of states of monoclinic \ce{Nb12O29} (PBE+U). Fermi level indicated by a dashed line. Up and down spins colored in red and blue, respectively. Flat and dispersive bands are present, with strong similarity to those in \ce{Nb22O54}.}
    \label{fig:mNb12O29_DFT+U_BS_DOS}
\end{figure}

\begin{figure}[!htb]
    \centering
    \includegraphics[scale=0.22]{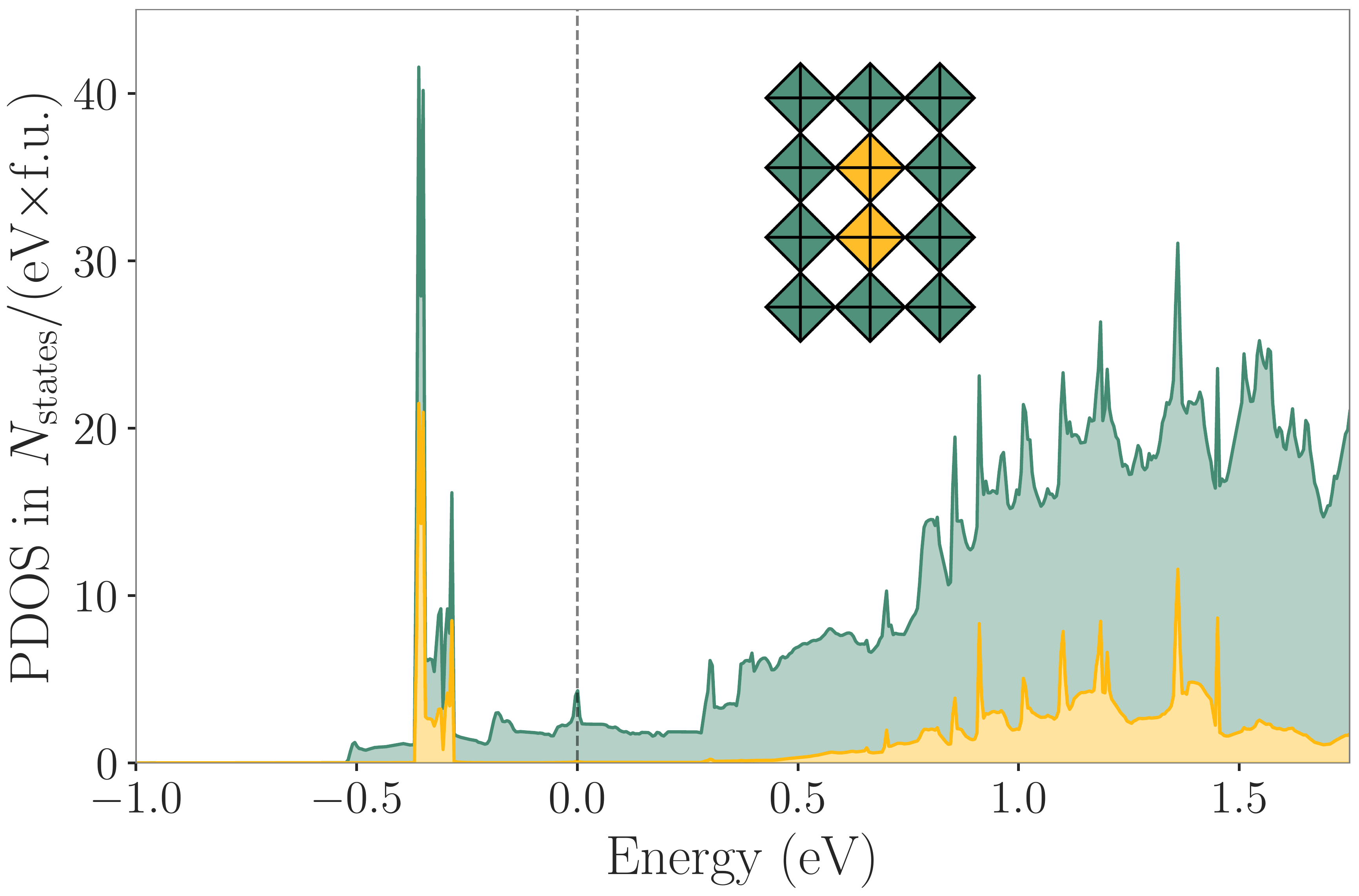}
    \caption{Projected density of states (PBE+U) for central (gold) and peripheral (green) niobium sites in monoclinic \ce{Nb12O29}. Central niobium sites contribute to the density of occupied states only in a narrow window that is associated with the flat bands.}
    \label{fig:mNb12O29_PDOS}
\end{figure}

\begin{figure}[!htb]
    \centering
    \includegraphics[scale=0.23]{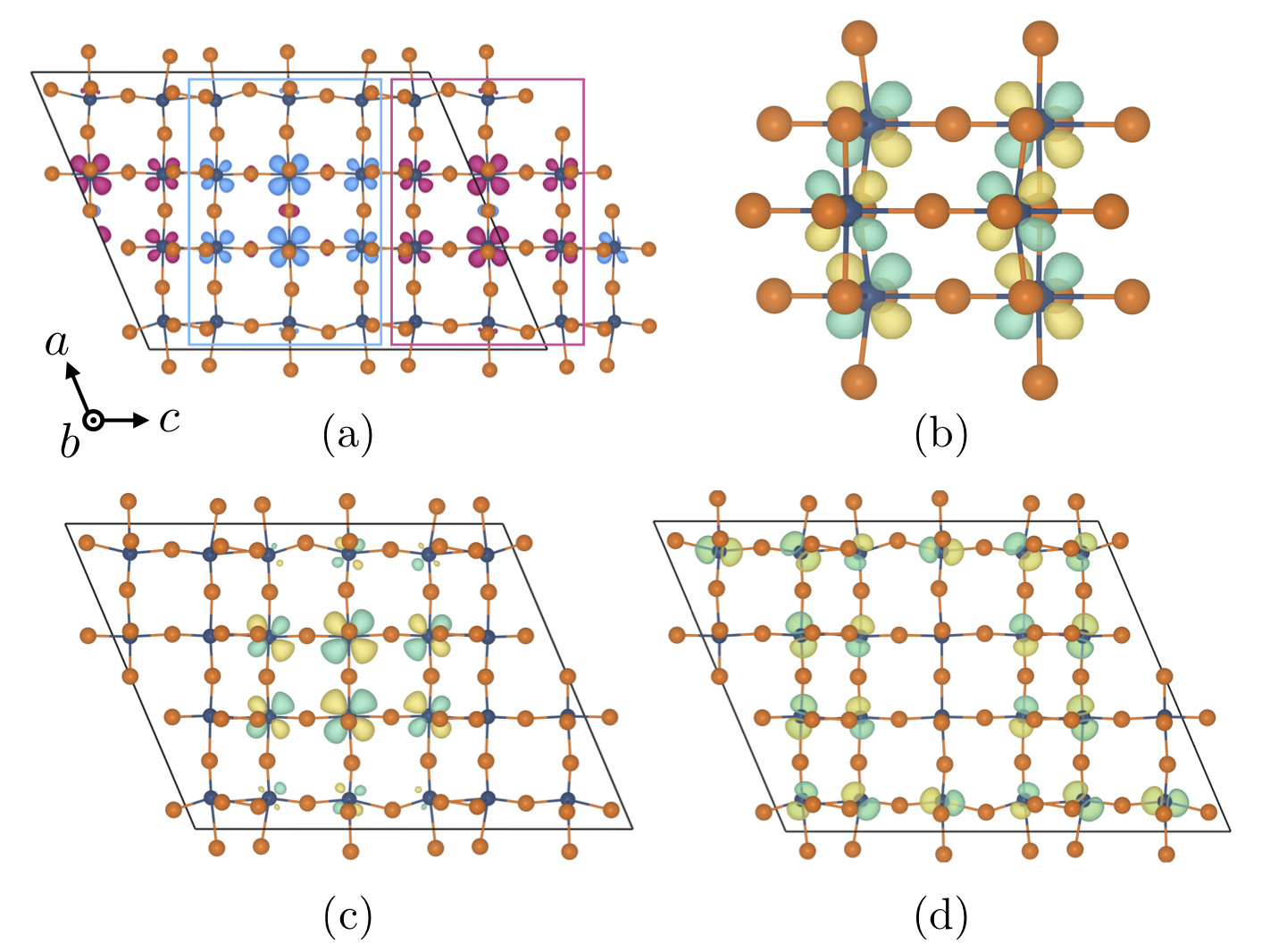}
    \caption{Spin density plot (a) and orbitals associated with localised (c) and delocalised (b,d) states in monoclinic \ce{Nb12O29}. Spin density (a) is predominantly located on the central niobium sites, and results from the occupation of localised states (c). Delocalised states have no contribution from the central niobium sites.}
    \label{fig:mNb12O29_wvfn}
\end{figure}

The bandstructure of monoclinic \ce{Nb12O29} shows two flat bands (one for each spin), which lead to two peaks in the DOS (Fig.~\ref{fig:mNb12O29_DFT+U_BS_DOS}). The shape of the real-space unit cell results in a Brillouin zone with two short and one long dimension, and the bandstructure path segments are similar to those in \ce{Nb22O54}. The bands for both spins lie exactly on top of each other due to the symmetry of the crystal structure, even though there is a spatial separation of spins (Fig.~\ref{fig:mNb12O29_wvfn}a). The flat bands coexist with more dispersive conduction bands, which show a dispersion which is largest in the $\mathbf{b}^{\ast}$ direction, making them effectively one-dimensional. Independent of the position of the flat bands, the larger number of electrons per block requires that some of the electrons fill dispersive conduction bands. This indicates a structural capacity for localised electrons. In \ce{Nb12O29} flat and dispersive bands are interspersed, while in \ce{Nb22O54}, the flat bands lie below the rest of the $d$-bands (cf. Fig.~\ref{fig:Nb22O54_DFT+U_BS}). Similar to \ce{Nb22O54}, the central Nb sites contribute exclusively to the occupied density of states in a narrow region that is associated with the flat bands (Fig.~\ref{fig:mNb12O29_PDOS}). The remainder of the conduction states involve contributions from the peripheral sites.

Kohn-Sham orbitals of the localised states (Fig.~\ref{fig:mNb12O29_wvfn}c) look remarkably similar to those in \ce{Nb22O54}, and are predominantly made up of Nb $d$-orbitals lying within the plane of the block. Electrons occupying these localised states are responsible for the non-zero spin density (Fig.~\ref{fig:mNb12O29_wvfn}a). Orbitals associated with dispersive bands (Fig.~\ref{fig:mNb12O29_wvfn}b,d) are made up of $d_{xy}$ and $d_{yz}$ atomic orbitals that are parallel to the crystallographic shear planes. The band dispersion along $\Gamma\to Z$ is explained by a reduction of in-phase overlap of the constituent atomic orbitals along the real-space $\mathbf{b}$ direction. The fact that the contributing atomic orbitals are parallel to the crystallographic shear planes and overlap face-on (Fig.~\ref{fig:mNb12O29_wvfn}b,d), rather than end-on, can be understood from a crystal field argument. For a transition metal ion in an ideal octahedral crystal field, the $t_{2g}$ orbitals form a degenerate set. The \ce{MO6} octahedra in shear phases, however, are far from ideal. When the degeneracy of the $t_{2g}$ orbitals is lifted by a distortion, those $d$-orbitals that do not overlap with any $\sigma$-type ligand orbitals will be lowest in energy and contribute to the low-energy $d$-bands.

\begin{figure}[!htb]
    \centering
    \includegraphics[scale=0.275]{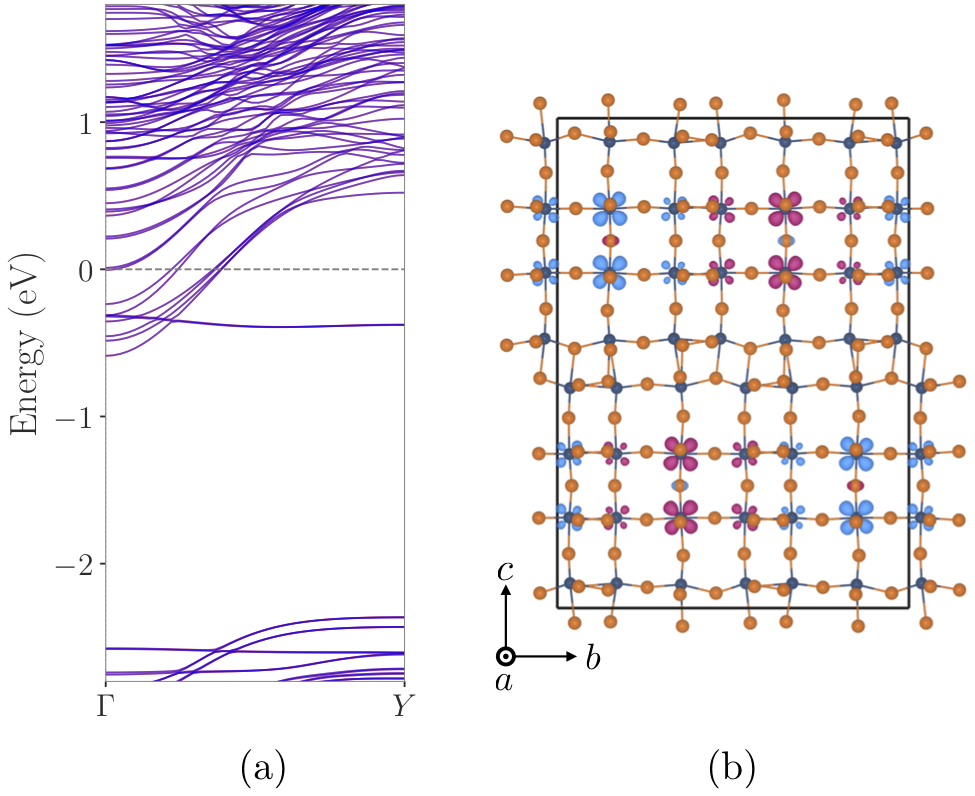}
    \caption{Bandstructure (PBE+U) (a) and spin density (b) of orthorhombic \ce{Nb12O29}. $Y=\mathbf{a}^{\ast}/2$. The orthorhombic and monoclinic \ce{Nb12O29} polymorphs show a strong similarity in their bandstructure and spin density distribution (cf. Figs. \ref{fig:mNb12O29_DFT+U_BS_DOS}, \ref{fig:mNb12O29_wvfn}a).}
    \label{fig:oNb12O29}
\end{figure}

The bandstructure of the orthorhombic \ce{Nb12O29} polymorph (Fig.~\ref{fig:oNb12O29}) looks very similar to that of the monoclinic polymorph. In fact, the results on the monoclinic polymorph presented above are transferable to the orthorhombic one. Experimentally, both polymorphs are metallic and show antiferromagnetic spin interactions, but only the monoclinic polymorph shows long-range magnetic order~\cite{andersen2005}. The magnetic susceptibility of \textit{m}-\ce{Nb12O29} can be fit using the Bonner-Fisher form, possibly indicating one-dimensional magnetism~\cite{lappas2002}. The differences are clearly subtle, and the small energy differences (10--20 K, around 1 meV) make comparisons using density-functional theory total energy differences difficult. However, the picture of the electronic structure of \ce{Nb12O29} that emerges is clear: for both polymorphs, conductivity and local moment magnetism are provided by different sets of electrons. Our conclusions on the orthorhombic polymorph are broadly in line with the first-principles study of Lee and Pickett~\cite{lee2015}. Those authors also found a coexistence of localised and delocalised electrons, with the localised spin residing in a large orbital dominated by the central niobium sites of the blocks, with delocalised electrons forming another subset. Our results as well as experimental studies using heat capacity measurements~\cite{cheng2009} and $\mu$SR spectroscopy~\cite{lappas2002} establish the presence of localised magnetic electrons in \textit{m}-\ce{Nb12O29}. We note that a previous study suggested the presence of itinerant moments in \textit{m}-\ce{Nb12O29} on the basis of GGA calculations~\cite{fang2014}. However, the high density of states at the Fermi level that was described to be the reason for the itinerant magnetism in fact arises from the flat band representing a localised state.

\subsection{\texorpdfstring{\ce{Nb25O62} and \ce{Nb47O116}}{ERROR}}

\begin{figure}[!htb]
    \centering
    \includegraphics[scale=0.225]{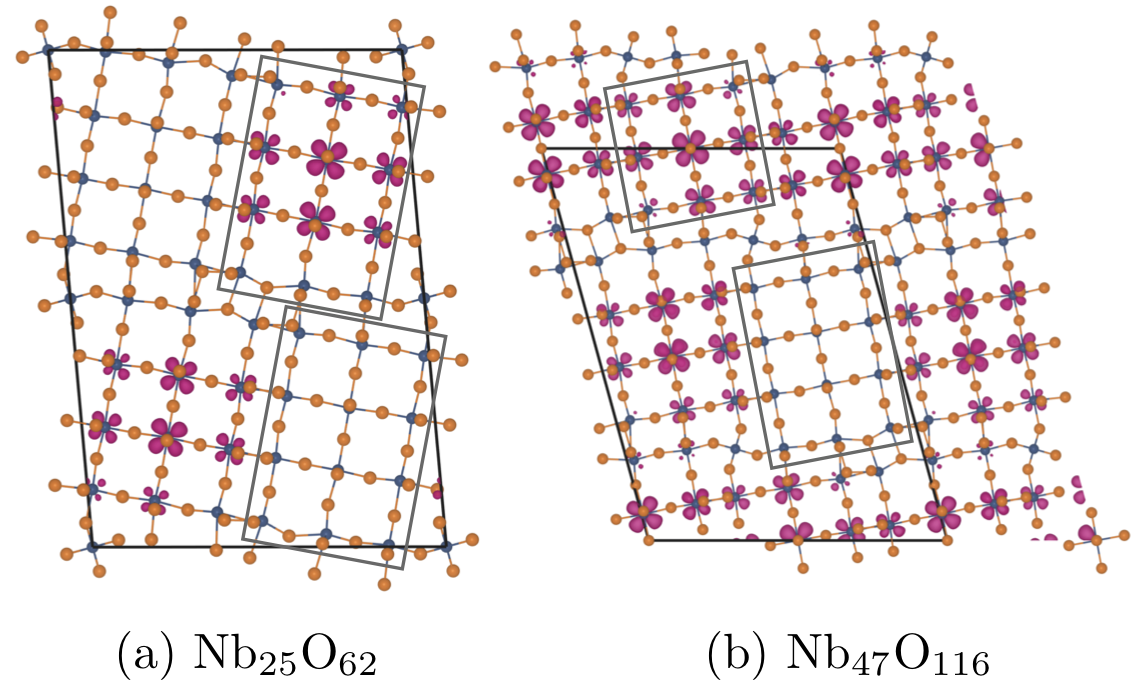}
    \caption{Summed charge densities from bands in (a) \ce{Nb25O62} and (b) \ce{Nb47O116}. Selected empty and filled localised states within blocks are framed by rectangles. The same (conventional) unit cell as in Fig. \ref{fig:xtalstrucs} is shown for \ce{Nb25O62}, but a smaller primitive cell for \ce{Nb47O116}.}
    \label{fig:Nb25O62_Nb47O116_Elec}
\end{figure}

The compounds \ce{Nb25O62} and \ce{Nb47O116} are less reduced than \ce{Nb22O54} and host less than one electron per structural block unit (Fig.~\ref{fig:xtalstrucs}, Table~\ref{tab:strucinfo}). The mutually occurring localised and delocalised electronic states that were found above for \ce{Nb22O54} and monoclinic \ce{Nb12O29} are also present in \ce{Nb25O62} and \ce{Nb47O116}. Localised states in blocks of the same size are nearly degenerate, and since only a fraction of the localised states is occupied (less than 1 electron per block), it is very difficult in a first-principles calculation to localise the electrons within a specific block. This could be done if the occupation of particular bands was constrained. Similarly, since the energy of the localised states depends on their occupation, judging the relative position of dispersive and localised states in these two compounds is very difficult. Charge densities for the localised states in the \ce{Nb47O116} and \ce{Nb25O62} are shown in Fig. \ref{fig:Nb25O62_Nb47O116_Elec}. It seems very likely that both \ce{Nb25O62} and \ce{Nb47O116} possess only localised electrons, occupying a fraction of these localised states. Since \ce{Nb47O116} is a unit cell level intergrowth of \ce{Nb22O54} and \ce{Nb25O62}, and \ce{Nb22O54} shows complete localisation of electrons, it is very likely that electrons should also fully localise in \ce{Nb47O116}, at least in those parts of the structure that derive from \ce{Nb22O54}.

\subsection{\texorpdfstring{H-\ce{Nb2O5}}{ERROR}}

H-\ce{Nb2O5} is the high-temperature phase of niobium pentoxide, and crystallises in space group $P2/m$ (Table \ref{tab:strucinfo}, Fig. \ref{fig:xtalstrucs}). As the parent compound of the crystallographic shear structures, its electronic structure provides a reference. However, since it is fully oxidised, all niobium ions have a $d^0$ configuration and there are no electrons occupying the conduction band.

\begin{figure}[!htb]
    \centering
    \includegraphics[scale=0.5]{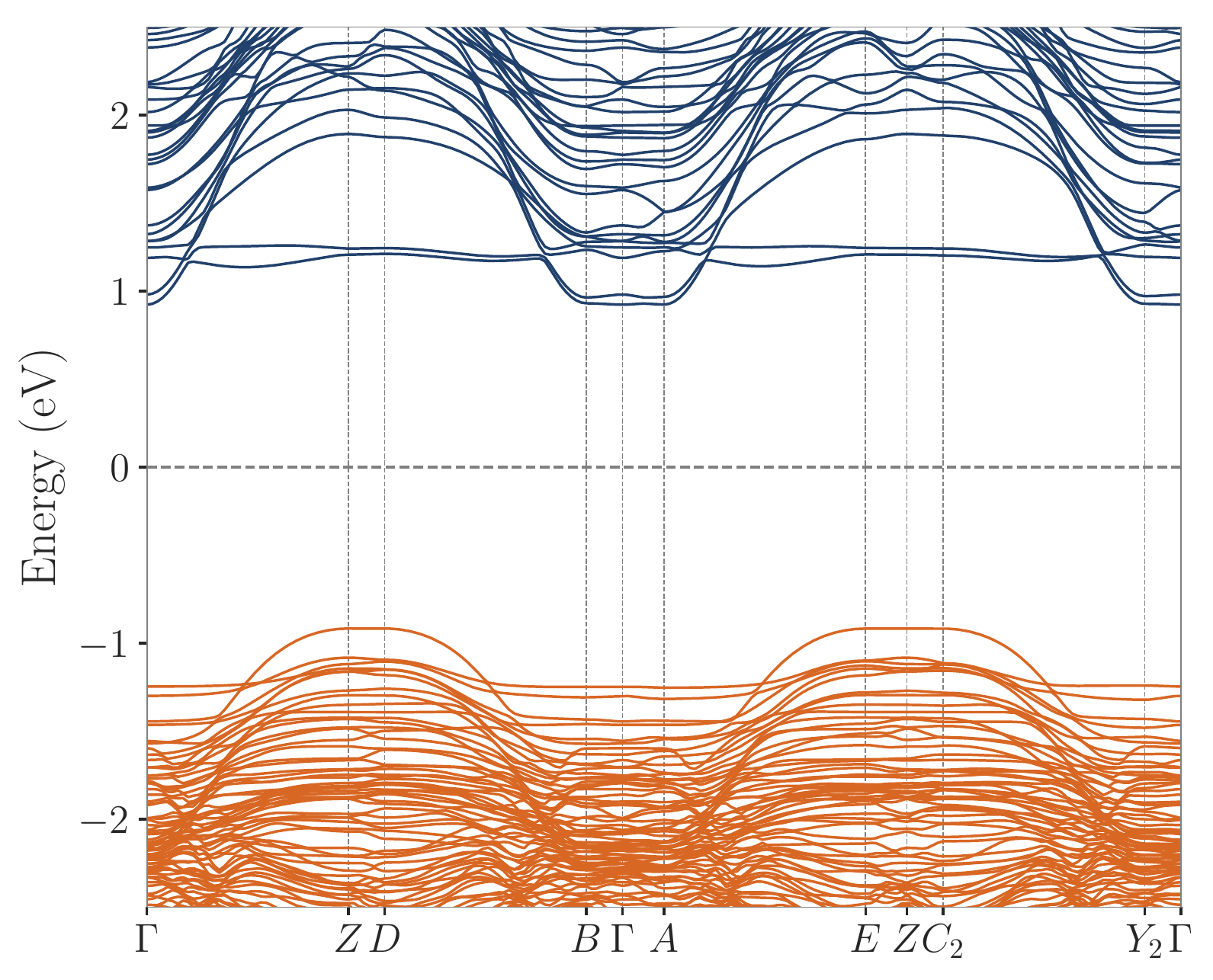}
    \caption{Bandstructure of H-\ce{Nb2O5} (PBE). Valence and conduction bands are colored in orange and blue, respectively. Flat and dispersive bands are present above the Fermi level (dashed line) similar to those in \ce{Nb22O54} and \ce{Nb12O29}, but are unoccupied.}
    \label{fig:HNb2O5_PBE_BS_DOS}
\end{figure}

\begin{figure}[!htb]
    \centering
    \includegraphics[scale=0.25]{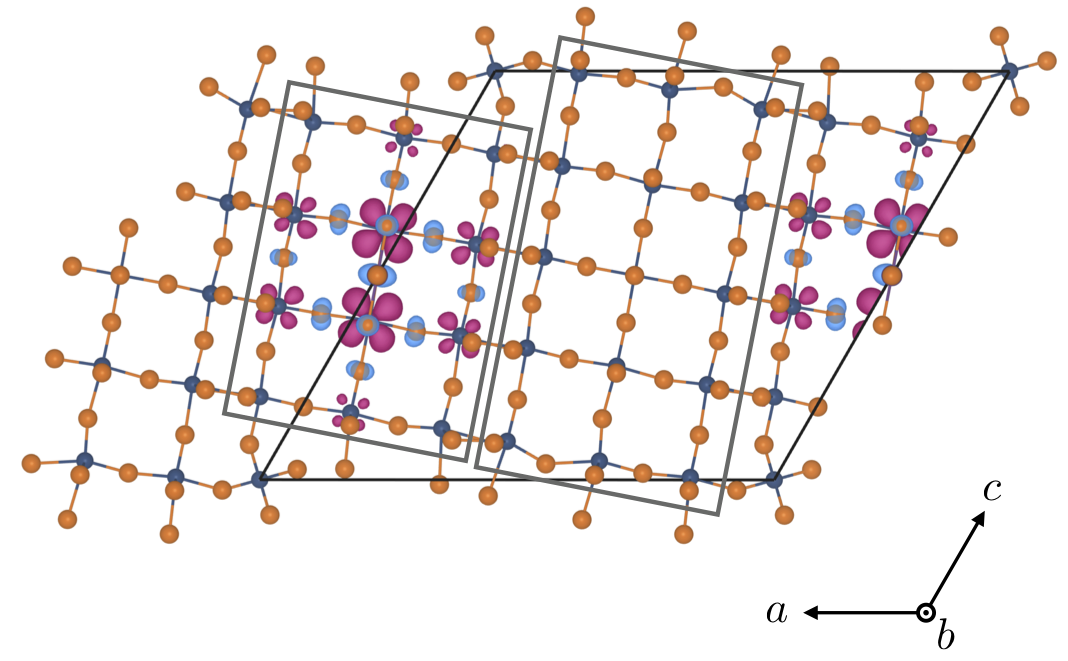}
    \caption{Spin density plot of lithiated H-\ce{Nb2O5}. A single lithium is located in the middle of the smaller block, inducing a localised state.}
    \label{fig:HNb2O5_Spindensity}
\end{figure}

The bandstructure of H-\ce{Nb2O5} shows the presence of flat and dispersive bands (Fig. \ref{fig:HNb2O5_PBE_BS_DOS}), similar to those found for the other shear phases above. However, the relative position of these bands depends on whether or not they are occupied. Doping by alkali metal ions is one way to introduce electrons into the conduction band, and in the particular case of H-\ce{Nb2O5} this has a practical relevance. Transition metal oxides in general, and the niobium-based oxide shear phases of this work, are used as electrodes within lithium-ion batteries. Like oxygen removal, lithium intercalation is a method to $n$-dope the material. Similar behaviour can often be observed from charge doping and ion insertion, for example in \ce{Na_xWO3}\cite{walkingshaw2004}. The H-\ce{Nb2O5} phase has been studied extensively for lithium-ion battery applications and it is closely related to other shear phases that have been examined for the same purpose~\cite{griffith2016,griffith2018}. Inserting a single lithium per unit cell into the middle of the $3\times 4$ block results in a localised state similar to those present in the niobium suboxides (Fig. \ref{fig:HNb2O5_Spindensity}). Note that the electron is entirely localised within the $3\times 4$ block, with the $3\times 5$ block remaining empty. Oxygen removal (as in the suboxides above) and lithium intercalation (examined here) clearly result in similar electronic structure features.

\section{Discussion}

Our results establish that the presence of defect-like flat bands and metallic conduction states is an innate feature of block-type structures. This coexistence arises due to the two different types of niobium sites present in the crystal structures; the central \ce{NbO6} octahedra are purely corner-shared, the distance between niobium atoms is larger and orbital overlap is reduced. This isolation results in localised electronic states, while along the crystallographic shear planes, where Nb-Nb distances are smaller and orbitals overlap more strongly, delocalised states are present. Each block can host one localised electron that is, rather unusually, spread over multiple niobium sites. This spread over multiple sites explains why single crystal X-ray diffraction studies on \ce{Nb22O54} and {\it o}-\ce{Nb12O29} do not show the presence of charge ordering~\cite{mcqueen2007}, despite the detection of localised electrons by magnetic measurements~\cite{cava1991a}. As the electronic structure features are ultimately a result of the blocks as structural units, the same principles are likely to apply to other crystallographic shear phases in the \ce{WO3}-\ce{Nb2O5} and \ce{TiO2}-\ce{Nb2O5} phase diagrams.

\begin{figure}[!hb]
    \centering
    \subfloat[$\nu\leq1$]{\includegraphics[scale=0.22]{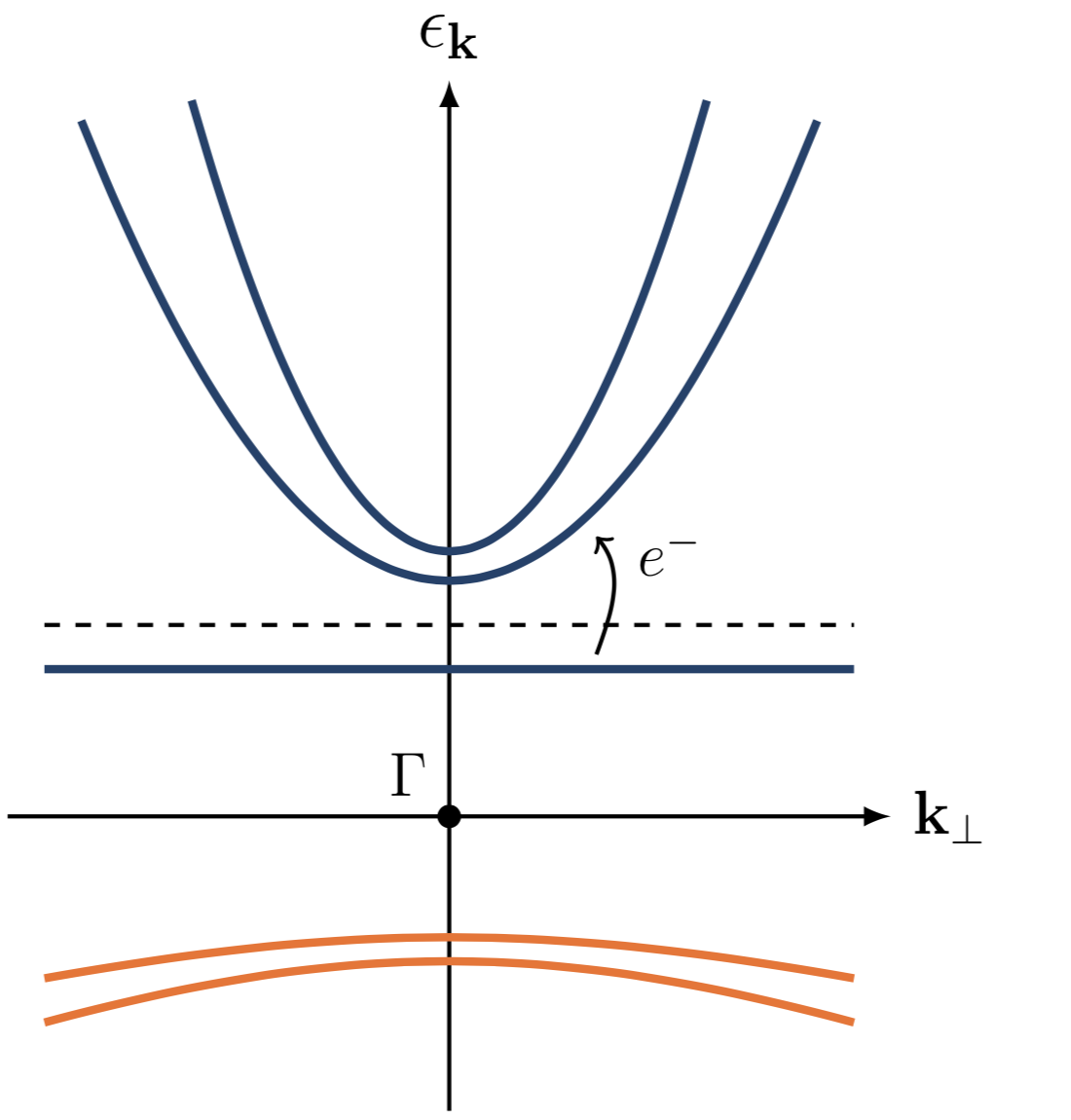}}
    \subfloat[$\nu>1$]{\includegraphics[scale=0.22]{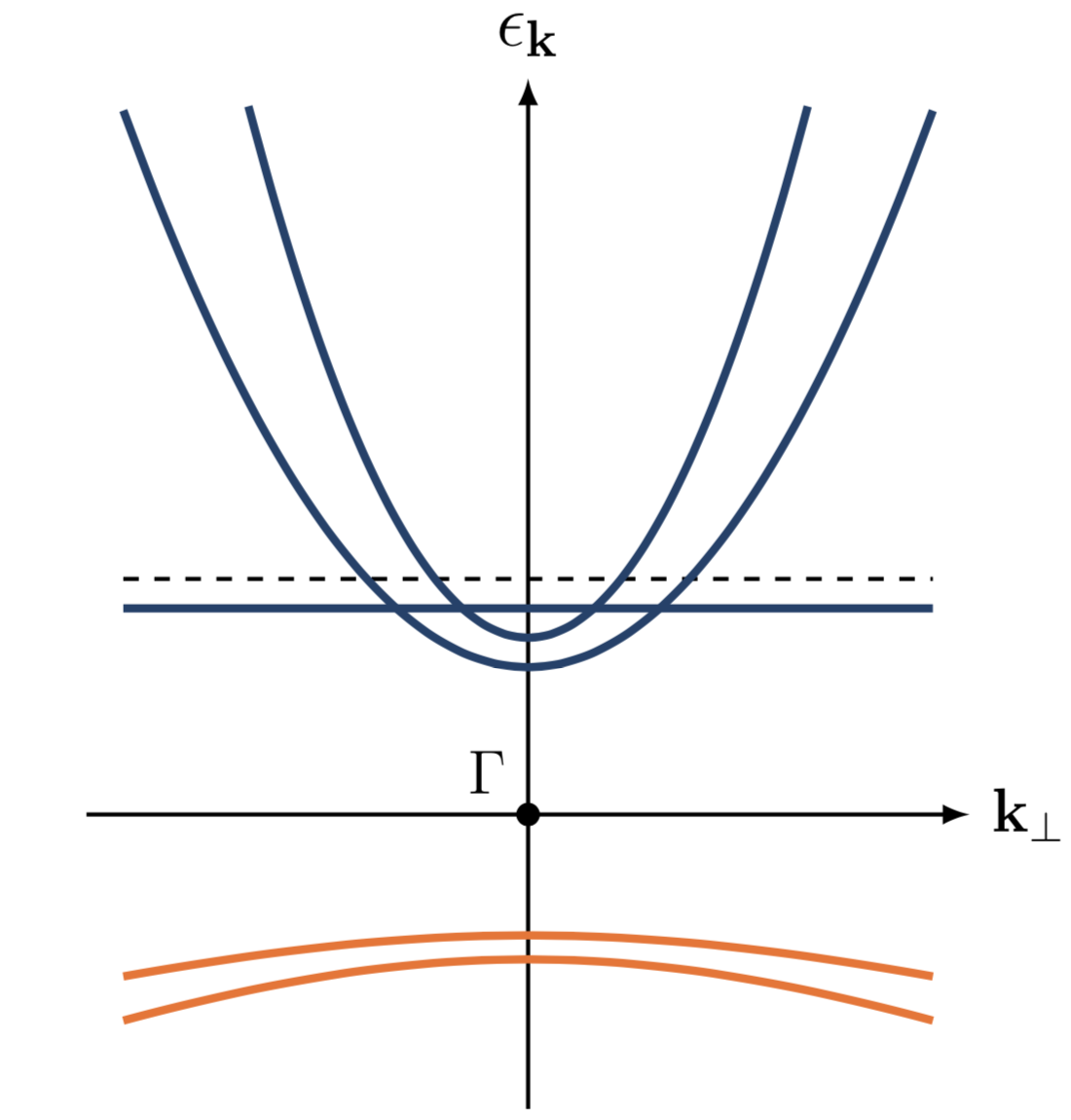}}
    \caption{Schematic of bandstructure for (a) filling fraction $\nu\leq$ 1 $e^{-}$/block and (b) $\nu>$ 1 $e^{-}$/block. O $2p$ and Nb $4d$ dominated bands are colored in orange and blue, respectively. Fermi level is indicated by a dashed line, $\mathbf{k}_{\perp}$ designates reciprocal space vector associated with the real space direction perpendicular to the block plane. The relative position of flat and dispersive bands changes with the filling fraction $\nu$.}
    \label{fig:Nb2O5-x_Scheme}
\end{figure}

\begin{figure*}
    \centering
    \subfloat[]{\includegraphics[scale=0.325]{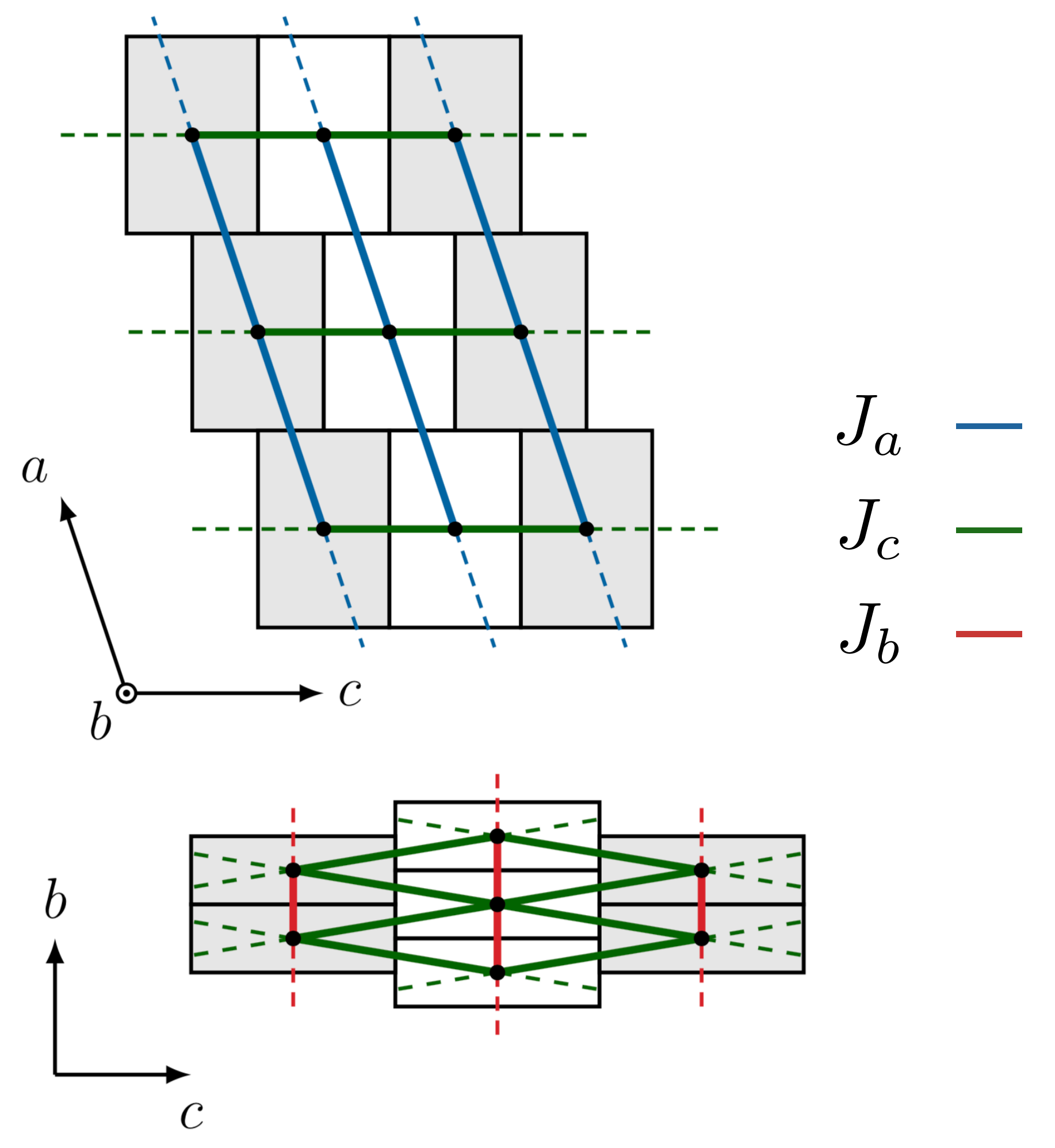}}
    \subfloat[]{\includegraphics[scale=0.25]{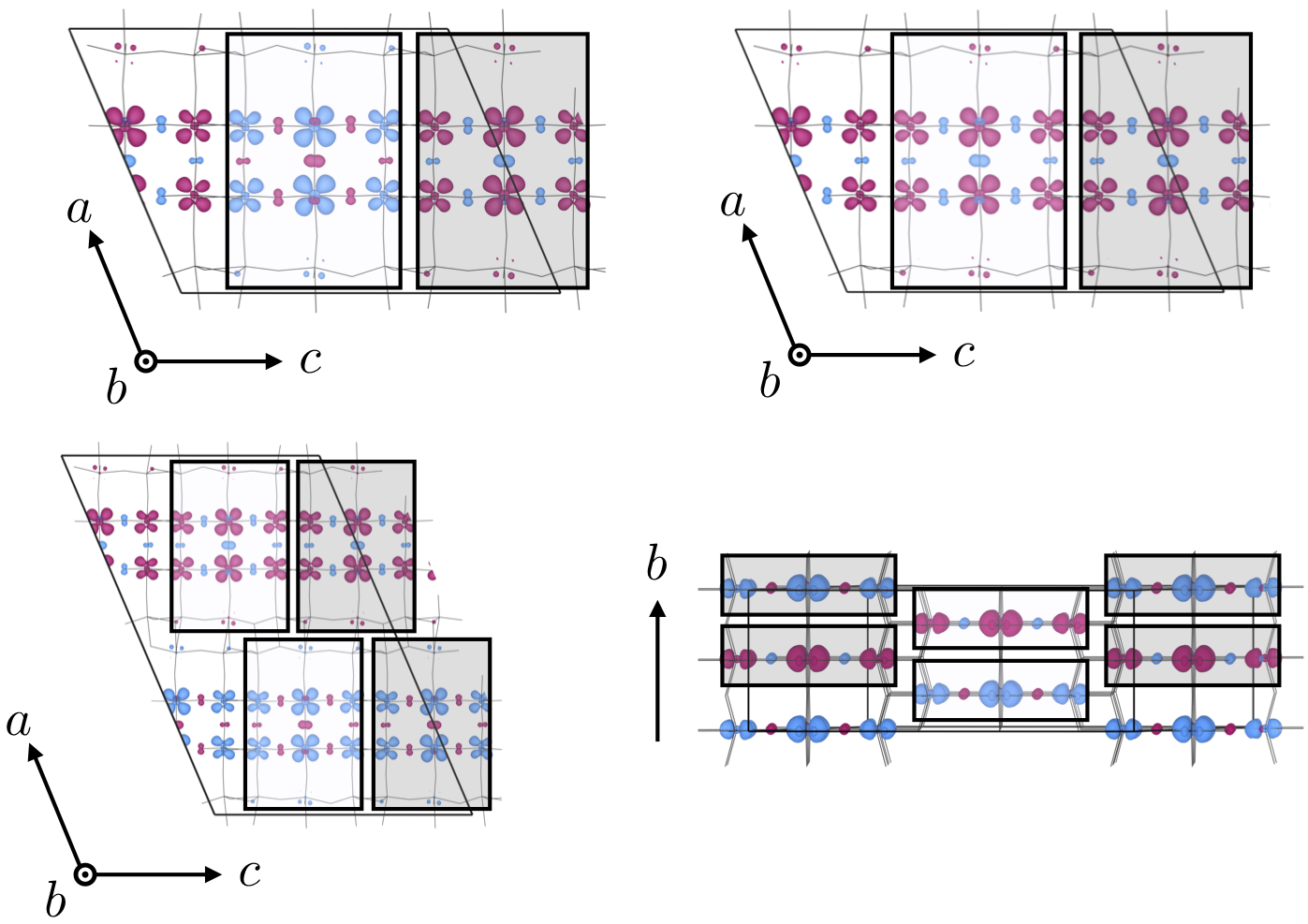}}
    \caption{a) Possible spin-spin interactions $J_i$ in monoclinic \ce{Nb12O29} along crystallographic $a$, $b$ and $c$ directions. b) Different spin arrangements in monoclinic \ce{Nb12O29}. AFM along $c$ (top left), fully FM (top right), AFM along $a$ (bottom left), AFM along $b$ (bottom right). White and grey blocks are offset by 0.5 $b$ throughout.}
    \label{fig:mNb12O29_Magnetism}
\end{figure*}

Regarding possible electronic conduction mechanisms in the niobium suboxides, the relevant quantities are the filling fraction $\nu$ (number of $e^-$ per block), and the energy gap between the flat and dispersive bands. For filling fractions of less than one, only localised states are filled (\ce{Nb25O62} and \ce{Nb47O116}, Fig.~\ref{fig:Nb25O62_Nb47O116_Elec}) and electrons can hop from one filled block to another empty one by a polaron hopping mechanism. The hopping process will have an activation energy. However, in \ce{Nb22O54} all localised states are filled and this hopping mechanism becomes impossible. With no metallic conduction electrons present, thermal excitation from the defect-like flat bands into the dispersive conduction bands might provide the dominant mechanism, as illustrated in Fig.~\ref{fig:Nb2O5-x_Scheme}a. This mechanism is reminiscent of doped semiconductors, and the activation energy associated with this process will depend on the separation between the flat and dispersive bands (cf. Figs~\ref{fig:Nb22O54_Theorycomp}, \ref{fig:Nb2O5-x_Scheme}a). Thermal excitation from flat into dispersive bands is also possible in \ce{Nb25O62} and \ce{Nb47O116}, and could coexist with a polaron hopping mechanism. Finally, in the case of \ce{Nb12O29}, all localised states are filled, but metallic conduction states are also partially filled (Fig.~\ref{fig:Nb2O5-x_Scheme}b). The result is metallic conductivity, which does not require thermal activation. Both R\"uscher et al.~\cite{ruscher1988} and Cava et al.~\cite{cava1991,cava1991a} have studied the conductivity of \ce{Nb2O_{5-$\delta$}} compounds and observed that all phases except \ce{Nb12O29} exhibit thermally activated conductivity. R\"uscher et al. also noted the effectively one-dimensional electron transport properties along the block columns~\cite{ruscher1992}, which are consistent with the calculated band dispersions. In addition, the experiments of Cava et al. show semiconducting electronic behaviour for \ce{Nb25O62} from 4--300 K; \ce{Nb47O116} and \ce{Nb22O54} exhibit semiconducting (thermally activated) conductivity from 0--250 K and from 0--100 K, respectively~\cite{cava1991a}. Beyond those temperatures, there is a metallic-like range of temperatures in which conductivity decreases again. Given this change in the temperature dependence of the conductivity from semiconducting to metallic, the flat bands associated with localised electrons are likely shallow donor levels (Fig.~\ref{fig:Nb2O5-x_Scheme}). The complex interplay between electron localisation and delocalisation in the suboxides is more similar to phenomena occurring in semiconductors on $n$-type doping, but distinctly different from metal-insulator transitions in transition metal oxides~\cite{imada1998}. Our results suggest that a similar phenomenon of crossover from localised to metallic conduction could occur on lithium doping of H-\ce{Nb2O5}, which might be observed with electrochemical, spectroscopic, or magnetic measurements.

Magnetic susceptibility measurements on the suboxides show that the number of localised moments increases with the degree of reduction~\cite{ruscher1991,cava1991a}. However, under the assumption that $g=2$, the number of moments calculated from the measurements is smaller than the number of introduced electrons~\cite{ruscher1991,cava1991a}. For \ce{Nb12O29}, this is consistent with one delocalised and one localised electron, but for the remaining suboxides this apparent reduction in the number of local moments is unexpected. Our first-principles results are consistent with complete localisation of electrons until a threshold is exceeded, and therefore all moments should be seen. Deviations in the $g$ value of the electrons might explain the discrepancy. As the electrons in these suboxides are well-localised, electron paramagnetic resonance (EPR) spectroscopy could provide some insight into the nature of the electronic states and the $g$ values. For \ce{Nb22O54} in particular, the different shapes of the magnetic orbitals could be used to detect electrons occupying specific blocks. Another possibility is that localised electrons contributing magnetic moments coexist with magnetically inactive electrons in all suboxides, not just \ce{Nb12O29}. However, we see no evidence for this in our calculations, and the thermally activated conductivity of \ce{Nb22O54}, \ce{Nb47O116} and \ce{Nb25O62} seems inconsistent with the presence of magnetically inactive (Pauli-paramagnetic) metallic electrons.

Long-range antiferromagnetic order is observed only in the monoclinic \ce{Nb12O29} phase below 12 K, all other niobium suboxides are paramagnetic~\cite{cava1991,cava1991a}. The Curie--Weiss constants of \ce{Nb2O_{5-$\delta$}} are in the range of \SIrange{0}{24}{\kelvin} (\SIrange{0}{2}{\milli\electronvolt}), and indicate antiferromagnetic interactions that become stronger with increasing degree of reduction~\cite{cava1991a}. In \ce{Nb25O62} and \ce{Nb47O116}, some of the localised states are empty (cf. Fig.~\ref{fig:Nb25O62_Nb47O116_Elec}), and the magnetic lattice is not fully filled. Independent of the strength of interaction, if not all spins have neighbours to couple with, or there is some randomness in the distribution of the spins, long-range magnetic order is unlikely to emerge. With first-principles calculations it is very difficult to address the question of why only {\it m}-\ce{Nb12O29} orders, but \ce{Nb22O54} and {\it o}-\ce{Nb12O29} do not, since the energy differences between different magnetic states are very small. However, we can discuss the possible spin-spin interactions simply based on the shape and orientation of the magnetic orbital within the crystal structure. We will focus in particular on monoclinic \ce{Nb12O29}, but similar considerations apply to the other suboxides. The magnetic orbital lies within the plane of the block. The two closest distances (two neighbours) between spins (two nearest neighbours) are along the block columns, with a separation of about 3.8 \si{\angstrom} ($J_b$, Fig.~\ref{fig:mNb12O29_Magnetism}a bottom). By symmetry, the interaction with spins in the four next-nearest neighbouring blocks along $c$, that are offset by 0.5 $b$, has to be the same ($J_c$, distance 10.6 \si{\angstrom}, Fig.~\ref{fig:mNb12O29_Magnetism}a bottom). In addition to that, each block is connected to two blocks on the same level in monoclinic \ce{Nb12O29} along the $a$ direction ($J_a$, distance 15.9 \si{\angstrom}, Fig.~\ref{fig:mNb12O29_Magnetism}a top), and four others offset by 0.5 $b$ along $a$ (distance 15.3 \si{\angstrom}, Fig.~\ref{fig:mNb12O29_Magnetism}a top). Different spin arrangements are easily obtained from DFT calculations (spin densities are shown in Fig.~\ref{fig:mNb12O29_Magnetism}b), but the energy differences between them are very small (few meV), and change significantly with the level of theory (PBE or PBE+U). Energy differences of a few meV are consistent with the interaction strengths obtained experimentally. The lowest energy magnetic ordering found in our calculations is antiferromagnetic along the $c$ direction (Fig.~\ref{fig:mNb12O29_Magnetism}b, top left).

\section{Conclusion}
We have shown that the electronic structure features common to $n$-doped crystallographic shear phases include (1)~effectively one-dimensional flat and dispersive bands corresponding to localised and delocalised electronic states (2)~electron localisation in orbitals spanning the block planes, and (3)~the partition of localised and delocalised states between central and peripheral niobium sites. Structural block units are also present in \ce{WO3}-\ce{Nb2O5}~\cite{roth1965} and \ce{TiO2}-\ce{Nb2O5}~\cite{wadsley1961,wadsley1961a} phases, and many of these mixed-metal shear phases have been explored as lithium-ion battery electrodes~\cite{griffith2018,guo2014}. The principles laid out in this work are likely transferable to these compounds, and are important for the interpretation of spectroscopic and electrochemical data.

The niobium suboxides show a transition from localised to delocalised electrons, but it is much smoother than commonly observed for metal-insulator transitions in transition metal oxides. In fact, our results portray the suboxides to be closer to $n$-doped semiconductors, but with a limited capacity for localised electrons. Once a filling threshold is exceeded, delocalised metallic electrons are simply added to existing localised electrons. This process is likely to occur  in heavily lithium-doped shear phases during battery operation. Similarly, the experimentally observed crossover from localised to delocalised electronic behaviour in \ce{WO_{3-x}}~\cite{salje1984} might have the same underlying mechanism, as \ce{WO_{3-x}} phases also exhibit some amount of crystallographic shear. More broadly, the niobium suboxides are an elegant example of the interplay between crystal and electronic structure, and the balance between electron localisation and delocalisation in oxides of an early transition metal.

\begin{acknowledgments}
The authors would like to thank Bartomeu Monserrat and Ieuan Seymour for useful discussions. We acknowledge the use of Athena at HPC Midlands+, which was funded by the EPSRC on grant EP/P020232/1, in this research via the EPSRC RAP call of spring 2018. C.P.K. thanks the Winton Programme for the Physics of Sustainability and EPSRC for financial support. K.J.G. thanks the Winston Churchill Foundation of the United States and the Herchel Smith Foundation. K.J.G. and C.P.G. also thank the EPSRC for funding under a programme grant (EP/M009521/1). The authors declare that the data supporting the findings of this study are available within the paper and its Supplementary Material files.
\end{acknowledgments}

\bibliography{refs}

\end{document}